\documentclass[american,aps,onecolumn,reprint,superscriptaddress]{revtex4-1}
\usepackage[T1]{fontenc}
\usepackage[latin9]{inputenc}
\usepackage{amsmath}
\usepackage{amsthm}
\usepackage{amssymb}
\usepackage{graphicx}

\usepackage[normalem]{ulem}
\makeatletter

\newtheoremstyle{customStyle1}  % name of the style to be used
{0pt}       % measure of space to leave above the theorem. E.g.: 3pt
{0pt}       % measure of space to leave below the theorem. E.g.: 3pt
{\normalfont}   % name of font to use in the body of the theorem
{\parindent}        % measure of space to indent
{\em}  % name of head font
{. --}   	 % punctuation between head and body
{.5em}       % space after theorem head
{\thmname{#1}\thmnumber{ #2}\thmnote{ (#3)}}  % Manually specify head

\theoremstyle{plain}

\newtheorem*{thm*}{\protect\theoremname}
\newtheorem*{prop*}{\protect\propositionname}
\newtheorem*{lem*}{\protect\lemmaname}
\newtheorem*{cor*}{\protect\corollaryname}

\ifx\proof\undefined
\newenvironment{proof}[1][\protect\proofname]{\par
	\normalfont\topsep6\p@\@plus6\p@\relax
	\trivlist
	\itemindent\parindent
	\item[\hskip\labelsep\scshape #1]\ignorespaces
}{%
	\endtrivlist\@endpefalse
}
\providecommand{\proofname}{Proof}
\fi

\usepackage{babel}
\usepackage{times}
\usepackage{pxfonts}
\usepackage{color}

\definecolor{mygrey}{gray}{0.35}
\definecolor{myblue}{rgb}{0.2,0.2,0.8}
\definecolor{myzard}{cmyk}{0,0,0.05,0}
\definecolor{mywhite}{rgb}{1,1,1}
\definecolor{myred}{rgb}{0.9,0.1,0.}
\definecolor{kangdaorange}{rgb}{1,0.4,0.}
\usepackage[colorlinks=true,citecolor=myblue,linkcolor=myblue,urlcolor=myblue]{hyperref}

\renewcommand{\vec}[1]{\text{\boldmath$#1$}}

\newcommand{\bra}[1]{\langle #1|}
\newcommand{\ket}[1]{|#1\rangle}

\newcommand{\ketbra}[2]{|#1\rangle\!\,\langle#2|}

\makeatother

\usepackage{babel}
\providecommand{\corollaryname}{Corollary}
\providecommand{\lemmaname}{Lemma}
\providecommand{\propositionname}{Proposition}
\providecommand{\theoremname}{Theorem}

\usepackage{etoolbox}
\usepackage[normalem]{ulem}
\newtoggle{showDeleted}
\toggletrue{showDeleted}
\iftoggle{showDeleted}{
	
	\newcommand{\comment}[1]{\textcolor{green}{#1}}
}{
	\newcommand{\streichen}[1]{}
	\newcommand{\comment}[1]{}
}

\begin{document}

\title{Detecting Non-Markovianity via Quantified Coherence: Theory and Experiments}
\date{\today}

\author{Kang-Da Wu}
\affiliation{CAS Key Laboratory of Quantum Information, University of Science and Technology of China, Hefei, 230026, People's Republic of China}
\affiliation{CAS Center For Excellence in Quantum Information and Quantum Physics, University of Science and Technology of China, Hefei, 230026, People's Republic of China}

\author{Zhibo Hou}
\affiliation{CAS Key Laboratory of Quantum Information, University of Science and Technology of China, Hefei, 230026, People's Republic of China}
\affiliation{CAS Center For Excellence in Quantum Information and Quantum Physics, University of Science and Technology of China, Hefei, 230026, People's Republic of China}

\author{Guo-Yong Xiang}
\email{gyxiang@ustc.edu.cn}
\affiliation{CAS Key Laboratory of Quantum Information, University of Science and Technology of China, Hefei, 230026, People's Republic of China}
\affiliation{CAS Center For Excellence in Quantum Information and Quantum Physics, University of Science and Technology of China, Hefei, 230026, People's Republic of China}

\author{Chuan-Feng Li}
\affiliation{CAS Key Laboratory of Quantum Information, University of Science and Technology of China, Hefei, 230026, People's Republic of China}
\affiliation{CAS Center For Excellence in Quantum Information and Quantum Physics, University of Science and Technology of China, Hefei, 230026, People's Republic of China}

\author{Guang-Can Guo}
\affiliation{CAS Key Laboratory of Quantum Information, University of Science and Technology of China, Hefei, 230026, People's Republic of China}
\affiliation{CAS Center For Excellence in Quantum Information and Quantum Physics, University of Science and Technology of China, Hefei, 230026, People's Republic of China}

\author{Daoyi Dong}
\affiliation{School of Engineering and Information Technology, University of New South Wales, Canberra, ACT, 2600, Australia}

\author{Franco Nori}
\affiliation{Theoretical Quantum Physics Laboratory, RIKEN Cluster for Pioneering Research, Wako-shi, Saitama 351-0198, Japan}
\affiliation{Physics Department, University of Michigan, Ann Arbor Michigan 48109-1040, USA}

\begin{abstract}
	The dynamics of open quantum systems and manipulation of quantum resources are both of fundamental interest in quantum physics. Here, we investigate the relation between quantum Markovianity and coherence, providing an effective way for detecting non-Markovianity based on the \textit{quantum-incoherent relative entropy of coherence} ($\mathcal{QI}$ REC). We theoretically show the relation between completely positive (CP) divisibility and the monotonic behavior of the $\mathcal{QI}$ REC. Also we implement an all-optical experiment to demonstrate that the behavior of the $\mathcal{QI}$ REC is coincident with the entanglement shared between the system and the ancilla for both Markovian and non-Markovian evolution; while other coherence-based non-Markovian information carriers violate monotonicity, even in Markovian processes. Moreover, we experimentally observe that non-Markovianity enhances the ability of creating coherence on an ancilla. This is the first experimental study of the relation between dynamical behavior of the $\mathcal{QI}$ REC and the phenomenon of information backflow. Moreover, our method for detecting non-Markovianity is applicable to general quantum evolutions.
\end{abstract}
\maketitle

\section{Introduction}
Quantum resource theory~\cite{ReversibleFrameworkforQuantumResourceTheories} studies the transformation and conversion of information under certain constrains, the quantification and manipulation of various resources are of central interest in quantum information, quantum thermodynamics, and other fields of physics~\cite{PhysRevAMarvian,marvian2014extending,PhysRevLettFernandoThermal,Thermodynamics2}. Recently, resource theories have inspired rigorous studies on the long-standing notions of non-classicality
in localized systems, where the development of coherence theory has become a fundamental task~\cite{streltsov2016quantum,QuantifyingCoherence,Giro14observable,MeasuringQuantumCoherencewithEntanglement,OperationalResourceTheoryofCoherence,Yuan15intrinsic}. 

Coherence is an intrinsically vulnerable resource, inevitably vanishing at macroscopic scales of space, time, and temperature~\cite{FrozenQuantumCoherence,TimeInvariantCoherenceNMR,huang2017non,qin2018dynamics,qin2018dynamics,lostaglio2017markovian,man2015cavity}. This becomes apparent in the study of the dynamical behavior of such resource in the presence of dissipation, where the system is rarely isolated and usually loses its information due to its environment~\cite{breuer2002theory,weiss2012quantum,PhysRevLettGeneraldynamics,PhysRevBquantumdots,PhysRevASpinNonMar,PhysRevAnonMarinputoutput,chen2015using,xiong2015non,pollock2018operational}. The problem of classifying memoryless dynamics and dynamics exhibiting memory effects has stimulated numerous investigations on the system-environment interaction. There are two main ideas: one idea, based on divisibility of the dynamical maps~\cite{gorini1976completely,lindblad1976generators}, is an analogy with the definition of classical stochastic processes; the other idea~\cite{MeasureBreuer2009} demonstrates that the memory effects may be accompanied with an information backflow, which is reflected by the non-monotonic behavior of some physical quantities~\cite{MeasureBreuer2009,EntanglementRivas2010,SunCPQFI2010,LuoShunlongQFI2010,Rajagopalfidelity2010,correlations2012Luo,bylicka2014non,Geometric2013Lorenzo,OperationalDivisibility2016Joonwoo,PRL2016nonMarTemporalsteering,strathearn2018efficient,PhysRevA.94.062126,xiong2017experimental,PhysRevA.92.062310}.

Rigorous studies on the dynamical behavior of quantified coherence in the presence of non-Markovian noise have recently attracted considerable attention~\cite{he2017non,radhakrishnan2017time,chanda2016delineating,mirafzali2019non,zhang2015role,coherencetrappingPRA,passos2018non,man2018temperature,ccakmak2017non,yu2018quantum,bhattacharya2016effect}. Moreover, information quantifiers based on coherence and the extended coherence with an ancilla have been proposed, for measuring the degree of non-Markovianity. It is known that coherence behaves monotonically in an incoherent Markovian evolution. However, for a general Markovian evolution, the dynamical behavior of quantum coherence is not necessarily monotonic. Moreover, the coherence-based quantifier that evolves monotonically in a certain basis may not evolve monotonically in another basis. Thus, two basic requirements for an advantageous coherence-based non-Markovianity measures are: (a) it is applicable for general evolutions; (b) for a Markovian quantum evolution, the monotonicity of the dynamical behavior of the quantifier is independent of the choice of reference basis.

In this work, we introduce a new way for detecting non-Markovianity based on the $\mathcal{QI}$ REC of a bipartite system~\cite{AssistedDistillationofQuantumCoherence,fanhuscireport2017,hu2016quantum}. Theoretically, we show that Markovianity implies the monotonic behavior of both the $\mathcal{QI}$ REC and the \textit{steering-induced coherence} (SIC, upper bounded by the $\mathcal{QI}$ REC)~\cite{fanhuscireport2017,hu2016quantum}. Experimentally, we compare our method with two existing approaches~\cite{he2017non,chanda2016delineating}, and verify the advantages of our new method for characterizing non-Markovianity. Moreover, we experimentally detect non-Markovianity via the non-monotonic behavior of both the $\mathcal{QI}$ REC and the SIC, which is coincident with previous results based on entanglement. Our work links the resource theory of coherence to quantum Markovianity.

\begin{figure}[htp!]
	\label{fig:theory}
	\includegraphics[scale=0.065]{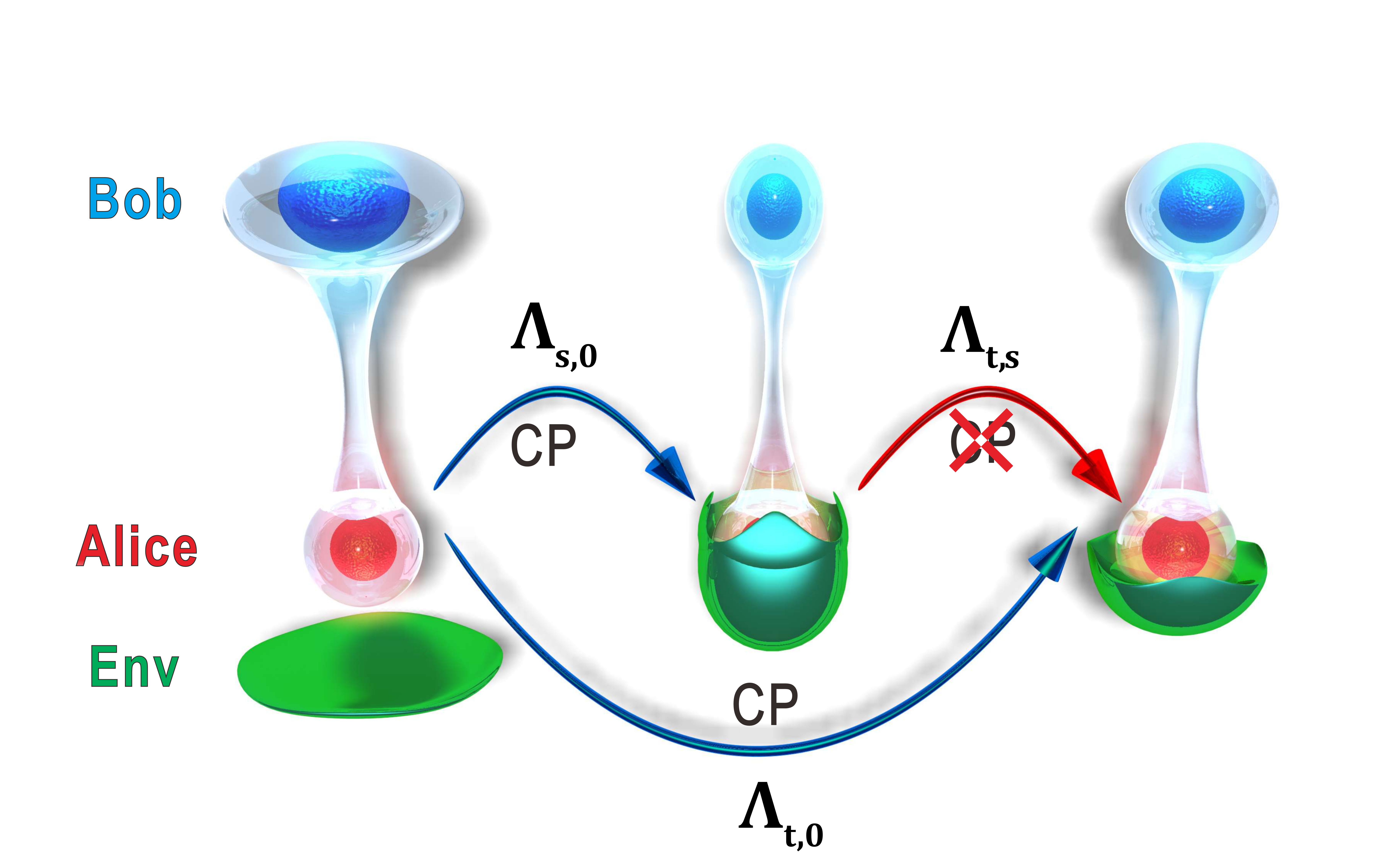}
	\caption{\label{fig:theory} \textbf{Theoretical Framework}. Consider a bipartite system involving Alice (red, system) and Bob (blue, ancilla) with nonzero initial $\mathcal{QI}$ REC, which is shown by the bigger volume of Bob. The environment is shown as green and can interact with Alice's system, while Bob is immune to the environment. Then Alice undergoes a quantum evolution which can be characterized by a family of $t$-parameterized dynamical maps $\{\Lambda_t\}$. If the evolution is CP divisible, then $\mathcal{QI}$ REC decreases monotonically. The CPTP map on Alice's system also affects the ability of preparing coherent states on Bob's system, in both the asymptotic limit and single-shot regimes (this is shown by the behavior of the SIC, and theoretically proved in the Supplementary Materials), which is shown by the decrease of Bob's volume. However, any temporal increase of the $\mathcal{QI}$ REC or the SIC indicates the violation of CPTP of the intermediate map $\Lambda_{t,s}$, and non-Markovianity.}
\end{figure}
\begin{figure*}[htp!]
	\label{fig:exp}
	\includegraphics[scale=0.064]{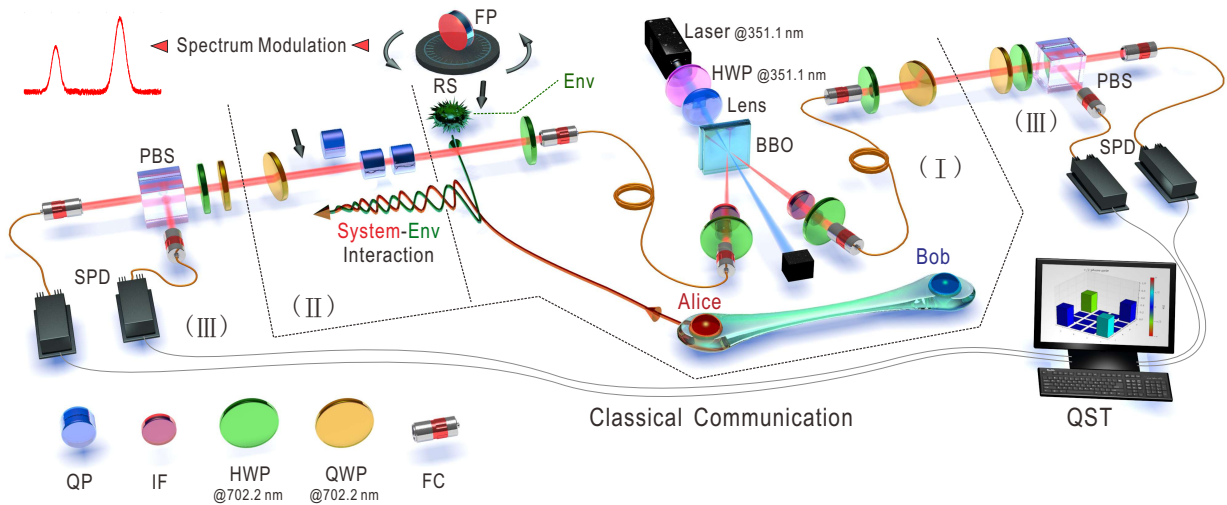}
	\caption{\label{fig:exp} The experimental setup is constructed by three modules: (\uppercase\expandafter{\romannumeral1}) state preparation, (\uppercase\expandafter{\romannumeral2}) evolution, and (\uppercase\expandafter{\romannumeral3}) detection. In (\uppercase\expandafter{\romannumeral1}), we can prepare two-photon states $\cos2\theta\ket{00}+\sin2\theta\ket{11}$, with arbitrary $\theta$ and $0\equiv H$ and $1\equiv V$. The environmental state can be modified by inserting a $\textmd{Fabry-P}\acute{\textmd{e}}\textmd{rot}$ cavity \cite{liu2011experimental}. In (\uppercase\expandafter{\romannumeral2}), the polarization and frequency degrees of freedom are coupled in a quartz plate (QP) in which different evolution times are realized by varying the thickness of the plates. The angles of the QPs can be controlled for dephasing in an arbitrary basis. Without modifying the spectrum of the frequency of Alice's photons, the environmental state can be modeled as a Gaussian distribution, resulting in a Markovian evolution of Alice's system. With the FP cavity (with thickness $h\approxeq0.06$ mm, and partial reflecting coating on each side) inserted, the environmental state of Alice can be modeled by the sum of two Gaussians centered at two different frequencies. In (\uppercase\expandafter{\romannumeral3}), the overall two-qubit quantum state $\rho^{AB}_t$ and a single-qubit state $\rho^A_t$ (on Alice's side) at different evolution times $t$ can be analyzed. Thus, we can obtain the experimental values for relevant information quantifiers. The coherence-based measures in different reference bases can be experimentally obtained for comparison. This setup can also be used for performing a local projective measurement on Alice and broadcasting the outcomes to Bob~\cite{wu2017experimentally}, which can experimentally detect the SIC. Keywords include: IF, interference filter; HWP, half-wave plate; QWP, quarter-wave plate; QP, quartz plate; FP, $\textmd{Fabry-P}\acute{\textmd{e}}\textmd{rot}$ cavity; BBO, $\beta$-barium borate; SPD, single photon detector; FC, fiber coupler; PBS, polarizing beam splitter; Env, environment; QST, quantum state tomography.}
\end{figure*}

\section{Theory}

First we describe the theoretical framework in an abstract fashion, including a brief introduction to quantum Markovianity, quantification of coherence, and its relation to non-Markovianity.

\textit{Quantum Markovianity}---In general, quantum evolution can be characterized by a family of one-parameter dynamical maps $\{\Lambda_{t,0}\}$ ($\Lambda_{t,0}$ is completely positive and trace preserving (CPTP) for any $t>0$, i.e., a legitimate quantum operation that maps the initial quantum state to the state at time $t$) and we assume that the inverse $\Lambda_{t,0}^{-1}$ exists for all time $t>0$. Thus, for any $t\,>\,s\,>\,0$, we can write the dynamical map for any $t$ into a composition
\begin{equation}\label{decomposition}
\Lambda_{t,0}=\Lambda_{t,s}\Lambda_{s,0}.
\end{equation}
However, even though $\Lambda_t^{-1}$ is well-defined and $\Lambda_{t,0}$, $\Lambda_{s,0}$ are completely positive (CP), the map $\Lambda_{t,s}$ does not need to be CP. If for any $t>s>0$, $\Lambda_{t,s}$ is CP, then the family of dynamical maps is said to be CP divisible. This leads to the definition of quantum Markovianity, providing a mathematical characterization of a map describing a memoryless evolution as a composition of physical maps. In this paper we adopt CP divisibility as the essential property of quantum Markovianity.

\begin{figure*}[htp!]
	\label{fig:data1}
	\includegraphics[scale=0.12]{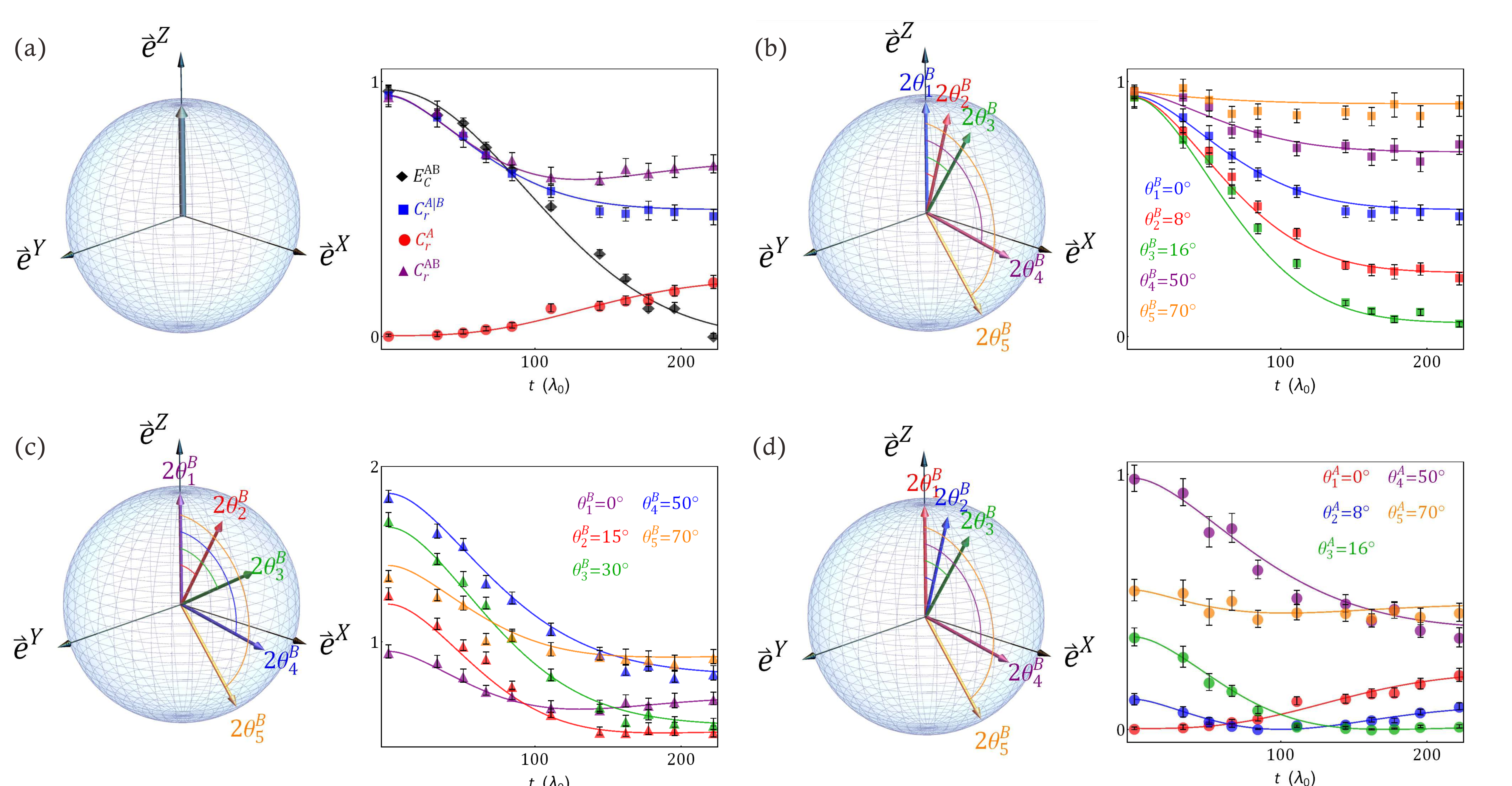}
	\caption{\label{fig:data1} \textbf{Experimental Results for monotonicity testing.} The Markovian evolution is constructed as decoherence in the eigenbasis of $\vec{\sigma}\cdot\vec{n}_0$, where $\vec{n}_0=\cos40^{\circ}\vec{e}^X+\sin40^{\circ}\vec{e}^Z$ and $\vec{\sigma}=(\sigma_x,\sigma_y,\sigma_z)$, by setting the rotation angle of the QPs to $20^{\circ}$. The evolution is implemented on both: a single system (Alice is prepared in $\ket{0}$) and a bipartite system (Alice and Bob initially share a maximally entangled state $\ket{\phi}^{AB}$). In (a), the experimental values for the concurrence, the $\mathcal{QI}$ REC, the extended coherence (with respect to the eigenbasis of $\sigma^A_z\otimes\sigma^B_z$), and the local coherence of Alice (with respect to the $\sigma^A_z$ basis) are shown as black diamonds, blue squares, purple up-triangles, and red disks, respectively. The dynamical behaviors of these coherence quantifiers are also tested with respect to different bases. The experimental values for the $\mathcal{QI}$ REC, the extended coherence, and the local coherence with respect to different bases are also obtained and shown in (b-d), respectively. For the $\mathcal{QI}$ REC, we choose five reference bases as the eigenbasis of $\sigma^A_z\otimes[\vec{\sigma}\cdot\vec{n}(\theta_i^B)]$, where $\vec{n}(\theta_i^B)=\sin2\theta_i^B\vec{e}^X+\cos2\theta_i^B\vec{e}^Z$, and $i=$1-5. The values of $\theta_i^B$ are shown in (b). In the right of (b), squares in different colors show the dynamical behavior of the $\mathcal{QI}$ REC in different bases. For the extended coherence, the different reference bases are also chosen as eigenbasis of $\sigma^A_z\otimes[\vec{\sigma}\cdot\vec{n}(\theta_i^B)]$ for various $\theta^B_i$. For the local coherence, five different bases of Alice are chosen according to the eigenbasis of $\vec{\sigma}\cdot\vec{n}(\theta_i^A)$. The choices of $\theta_i^{A(B)}$ are shown in the corresponding sub figures. All solid lines represent numerical simulations considering the fidelity of the states prepared in our laboratory, deduced assuming that the spectrum of Alice's photon is a Gaussian profile with a FWHM of 4 nm.}
\end{figure*}

\begin{figure}[htp!]
	\label{fig:data2}
	\includegraphics[scale=0.11]{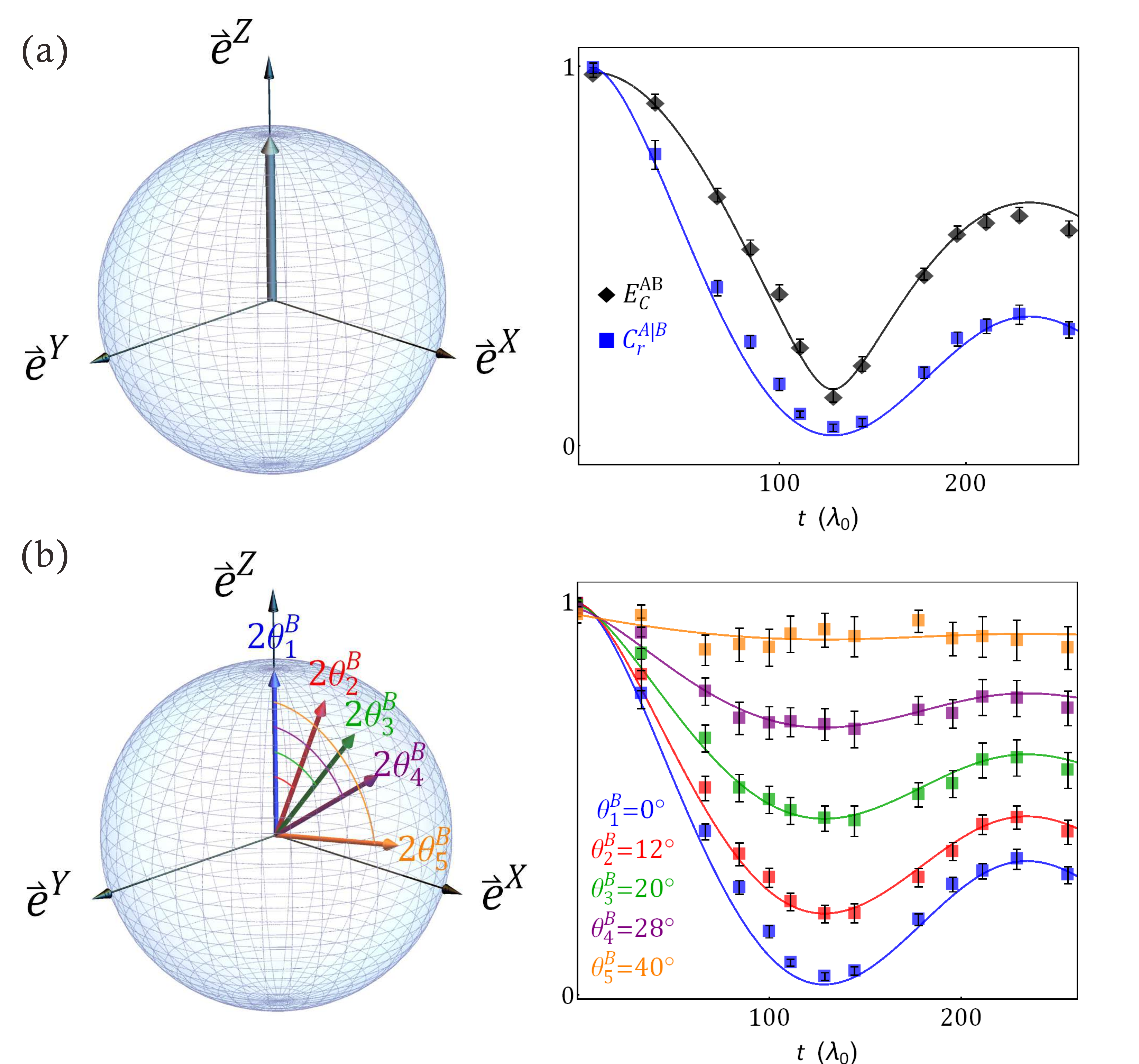}
	\caption{\label{fig:data2} \textbf{Experimental results for detecting non-Markovianity via $\mathcal{QI}$ REC.} The evolution is constructed as pure decoherence in the $\sigma_z$ basis. The experimental values for the concurrence and the $\mathcal{QI}$ REC (with respect to the eigenbasis of $\sigma^A_z\otimes\sigma_z^B$) at different evolution times $t$ are shown as black diamonds and blue squares in (a). For the $\mathcal{QI}$ REC, four different bases are chosen according to the eigenbasis of $\sigma_z^A\otimes[\vec{\sigma}\cdot\vec{n}(\theta_i^B)]$ with different $\theta_i^B$. All solid lines represent numerical simulations considering experimental imperfections, which are deduced assuming that the spectrum of Alice's photon is a sum of two Gaussians centered at two different frequencies, corresponding to wavelengths 700.6 nm and 703.3 nm with amplitudes 0.65 and 0.35.}
\end{figure}

\begin{figure}[htp!]
	\label{fig:data3}
	\includegraphics[scale=0.11]{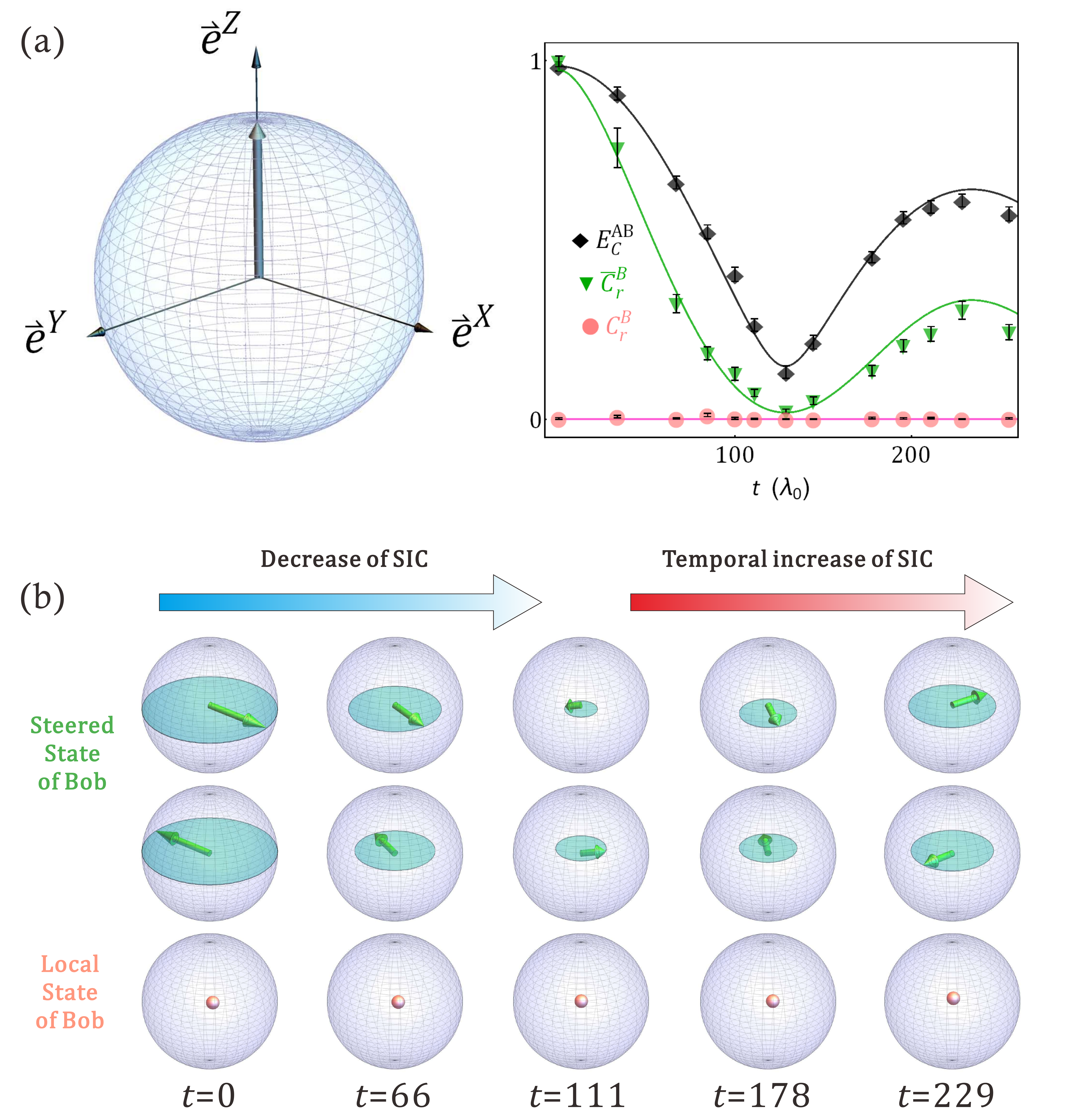}
	\caption{\label{fig:data3} \textbf{Experimental results for detecting Non-Markovianity via SIC.} The dynamical behavior of the SIC (in the $\sigma_z^B$ basis) was also investigated under the aforementioned evolution. The experimental values are shown in (a).  In the presence of the non-Markovian decoherence noise, the optimal measurement on Alice's system is $\sigma_x^A$ measurement on Alice. (b) illustrates the dynamical behavior of the SIC with respect to the $\sigma^B_z$ basis. The assisted state conversion process consists of a $\sigma_x^A$ measurement and the broadcast of outcomes $0$ or $1$ to Bob. Bob then can prepare a more coherent state at each time $t$. Bob's two steered states and local states are shown as Bloch vectors in green and pink respectively, and the radius of the disks $R(t)=\sqrt{r_x(t)^2+r_y(t)^2}$ show the coherence of each state. The experimental values for the local coherence on Bob's system are shown in (a) as pink disks.}
\end{figure}

\textit{Quantum-incoherent relative entropy of coherence}---In the resource theory of coherence, an orthogonal basis $\{\ket{i}\}$ is considered classical. Any mixture of such states is termed \textit{incoherent}. And \textit{incoherent operations} are naturally introduced as physical transformations that do not create coherence~\cite{QuantifyingCoherence}.

The amount of coherence for a general quantum state quantifies how close it is from the set of incoherent states with respect to a given reference basis, i.e., $C_r\equiv\mathcal\min_{\chi\in\mathcal{I}}S(\rho\|\mathcal{I})$, where $S(\rho\|\chi)=\textmd{Tr}[\rho\log_2\rho-\rho\log_2\chi]$ denotes the quantum relative entropy~\cite{vedral2002role}, and $\mathcal{I}$ denotes the set of incoherent states. 

Strictly, $C_r$ never increases under CPTP incoherent operations. However, when taking the advantage of assistance~\cite{AssistedDistillationofQuantumCoherence,streltsov2017towards}, one can increase $C_r$. Considering a bipartite system (Alice and Bob), assume Alice can perform any local projective measurement on her system and broadcast the outcomes to Bob. Then Bob can prepare more coherent states than his own states. In particular, the SIC \cite{fanhuscireport2017,hu2016quantum} is defined as the maximal average coherence on Bob's side, $\bar{C}_r^B(\rho^{AB})=\max_{\mathcal{M}_A}\sum_mp_mC_r(\rho^B_m),$ where the optimization is taken over all projective measurements $\mathcal{M}_A$ on Alice and $\rho_m$ is the state corresponding to the measurement outcome $m$. 

The SIC captures the steerability from Alice to Bob, where coherence on Bob's side is demanded. Generally, the identification of the SIC needs a non-trivial optimization over all possible measurements of Alice. The upper bound of this quantity, the $\mathcal{QI}$ REC was recently introduced in~\cite{AssistedDistillationofQuantumCoherence,streltsov2017towards},
\begin{equation}\label{QIREC}
C_r^{A|B}(\rho^{AB})\equiv\min_{\chi^{A|B}\in\mathcal{I}^{A|B}}S(\rho^{AB}\|\chi^{A|B}),
\end{equation}
where $\chi^{A|B}$ denotes the bipartite states that are \textit{quantum-incoherent} ($\mathcal{QI}$); i.e., $\chi^{A|B}=\sum_ip_i\sigma^A_i\otimes\ketbra{i}{i}^B$, where $\sigma^A_i$ represents an arbitrary quantum state, belonging to Alice.
Note that the $\mathcal{QI}$ REC is a measure that is different from entanglement or correlation measure (see the Supplementary Material for details on the $\mathcal{QI}$ REC).

\textit{Non-Markovianity witness via the $\mathcal{QI}$ REC}---The theoretical framework is shown in Fig.~\ref{fig:theory}. Considering the aforementioned bipartite system and assuming Alice will interact with her environment while Bob is kept isolated, then we have the following results.

\textit{Theorem} 1---In a bipartite state (Alice and Bob), assume that Alice undergoes an evolution characterized by $\{\Lambda_{t,0}\}$. Then the $\mathcal{QI}$ REC decreases monotonically if $\{\Lambda_{t,0}\}$ is CP divisible, i.e., for any $t>s>0$,
\begin{equation}\label{QIREC}
C_r^{A|B}(\rho^{AB}_{s})\geq C_r^{A|B}(\rho^{AB}_{t}),
\end{equation}
where $\rho^{AB}_{s}$ and $\rho^{AB}_{t}$ are the states after the evolution times $s$ and $t$, respectively. Moreover, the SIC decreases monotonically if $\{\Lambda_{t,0}\}$ is Markovian.

The result demonstrates that the $\mathcal{QI}$ REC on Bob's side provides a new method for characterizing the CP divisibility of a general quantum process on Alice's side. And the violation of monotonic behavior of either the $\mathcal{QI}$ REC or the SIC indicates non-Markovianity. The proof is presented in the Supplementary Material.

\section{Experiments}

In our experiments, the system-environment interaction is provided by the coupling of the polarization degree (Alice, the open system) and frequency degree (environment) of Alice's photons. The experimental setup is illustrated in Fig.~\ref{fig:exp}, and is constructed by three modules (for more information, see the Supplementary Materials). 

Our experiments contain two parts. In the first part, we implement a monotonicity test. In particular, we rotate the angles of all QPs to $20^{\circ}$, producting a pure decoherence in the eigenbasis of $\vec{\sigma}\cdot\vec{n}_0$ on Alice, where $\vec{n}_0=\cos2\theta\vec{e}^X+\sin2\theta\vec{e}^Z$ and $\theta=20^{\circ}$, maximizing the non-monotonic behavior of the extended coherence and the local coherence. We prepare both a Bell state $\ket{\psi}^{AB}$ (shared by Alice and Bob) and an incoherent pure state $\ket{0}^A$. Then we obtain the concurrence $E^{AB}_c(\rho^{AB}_t)$, the $\mathcal{QI}$ REC, the extended coherence, and the local coherence at various evolution times $t$. The experimental values for the above quantifiers are shown as black diamonds, blue squares, purple up-triangles and red disks in Fig.~\ref{fig:data1}(a), with respect to the eigenbasis of $\sigma^A_z\otimes\sigma^B_z$. The experimental dynamics for the $\mathcal{QI}$ REC, the extended coherence and the local coherence with respect to different bases are also shown in Figs.~\ref{fig:data1}(b-d), respectively. From these results, we can see that both the behaviors of the extended coherence and the local coherence are non-monotonic and basis-dependent; only the $\mathcal{QI}$ REC behaves monotonically during the Markovian process.

In the second part of the experiment, we implement a quantum evolution that is non-Markovian, for exploring the behaviors of the $\mathcal{QI}$ REC and the SIC in the presence of non-Markovian system-environment interactions. Since either the extended coherence or the local coherence behaves non-monotonically even in a Markovian process, we do not show them in the second part of the experiments. The evolution is constructed as decoherence in the $\sigma_z$ basis. We insert a $\textmd{Fabry-P}\acute{\textmd{e}}\textmd{rot}$ cavity into Alice's path, resulting in a modification of the spectrum of the frequency of Alice's photons. At different evolution times $t$, the values for concurrence and the $\mathcal{QI}$ REC (with respect to different bases) are obtained. The relevant experimental values are shown as black diamonds and blue squares in Figs.~\ref{fig:data2}(a, b). The dynamical behaviors of the SIC and the local coherence of Bob $C_r^B(\rho^{B}_t)$ are also investigated (in the $\sigma^B_z$ basis) and shown as green down-triangles and pink disks in Fig.~\ref{fig:data3}(a). And the evolution of the optimal steered states are shown in Fig.~\ref{fig:data3}(b). It is clear that the non-Markovianity can be captured by both the $\mathcal{QI}$ REC and the SIC.

All solid lines represent numerical simulations considering the experimental imperfections. In the Supplementary Material, we show that the experimental results admit a theoretical analysis, and we also perform numerical simulations to show that both the $\mathcal{QI}$ REC and the SIC are applicable to a large range of quantum processes for witnessing non-Markovianity.

\section{Discussion}

In this work, we theoretically provided a method for characterizing Markovianity based on the $\mathcal{QI}$ REC (between an open system and an ancilla) and experimentally investigated the evolution of the local coherence, the extended coherence, the $\mathcal{QI}$ REC, and the SIC in \emph{both} Markovian and non-Markovian processes. We highlight the two-fold advantages of our method: on the one hand, it overcomes the constraint in~\cite{he2017non,passos2018non,chanda2016delineating} as it can be applied to general quantum evolutions; on the other hand, the information carrier based on $\mathcal{QI}$ REC does not need any non-trival optimization. Moreover, in the experiments with the non-Markovian processes, these results show that the information backflow can enhance the ability of preparing coherent states on the ancilla system, in both asymptotic and single-shot settings, linking non-Markovianity to quantum resource theory. 

A full understanding of these connections still remains open. Quantum coherence has been regarded as a type of resource which is more fundamental than quantum correlations, and current researches highlight the links between different quantum resources~\cite{streltsov2015measuring,ConvertingCoherencetoQuantumCorrelations,PRLresourceonversion}. Though relatively elegant solutions come up in the context of coherence and correlations, a natural question arises that whether there exists a construction for the links between general quantum resources and non-Markovianity. 

Another promising line of research is studying the dynamics of resource conversion processes. It has been put forward that utilizing correlated resources, together with measurement feed-forward, can be advantageous in implementing certain gates in measurement-based computation~\cite{RausB01}. If the non-Markovianity could be related to enhancing the operational advantages of such protocols, it could imply that operational benefits can emerge when information backflow takes place.

The work at USTC is supported by the National Natural Science Foundation of China under Grants (Nos. 11574291, 11774334, and 61828303), the National Key Research and Development Program of China (No.2017YFA0304100), Key Research Program of Frontier Sciences, CAS (No.QYZDY-SSW-SLH003), National Key R \& D Program (2016YFA0301700), and
Anhui Initiative in Quantum Information Technologies. D.D. acknowledges partial support by the Australian Research Council's Discovery Projects Funding Scheme under Project DP190101566.

\begin{widetext}
\appendix	
	
	\section{\label{sec:appendix1}Theoretical tools}
	A \textit{quantum resource theory} (QRT) has several indispensable ingredients, including: constraints, states which contain no resource, and the measure for how much resource that a state possesses. The constraints, known as free operations, in a QRT are often desirable from a practical perspective that reflects current experimental capabilities. The states that contain no resource are often referred to as free states, and these can be generated by free operations without any cost. In the following, let us first briefly recall some basic information about free states and free operations in coherence theory.
	
	\subsection{\label{appendix1:sub1}Free states in the resource theory of quantum coherence}
	The free states in the resource theory of quantum coherence are incoherent states \cite{QuantifyingCoherence}. A quantum state $\rho$ is said to be incoherent in a given reference basis $\{\ket{i}\}$ if that state is diagonal in this basis, i.e.,
	\begin{equation}\label{incoherentstate}
	\rho=\sum_ip_i\ketbra{i}{i},
	\end{equation}
	where $0\,\leq\,p_i\,\leq\,1$ for all $i$, and $\sum_i p_i=1$. The reference basis is often chosen according to the context of the story, usually motivated by physical grounds of being easy to synthesize or store, e.g., eigenbasis of the Hamiltonian in quantum thermodynamics, polarization or path degree of a photon, internal states of an ionic atom, and so on.
	
	For bipartite systems partitioned by $A$ and $B$, each with respective local reference bases $\{\ket{i}^A\}$ and $\{\ket{j}^B\}$, the incoherent states take the form
	\begin{equation}\label{iistate}
	\chi^{AB}=\sum_{ij}p_{ij}\ketbra{i}{i}^A\otimes\ketbra{j}{j}^B,
	\end{equation}
	where $0\,\leq\,p_{ij}\,\leq\,1$, and $\sum_{ij}p_{ij}=1$. Note that in the aforementioned bipartite systems, we can also choose an orthogonal complete set of entangled pure states as the reference basis. In this case, the bipartite systems are viewed as a single physical system.
	
	In the above case, coherence in both $A$ and $B$ are viewed as resources. In the task of \textit{assisted distillation of quantum coherence}~\cite{AssistedDistillationofQuantumCoherence}, involving a bipartite system (Alice and Bob) where only the coherence of Bob is viewed as a resource, the \textit{quantum-incoherent} ($\mathcal{QI}$) states are introduced and can be regarded as free states,
	\begin{equation}\label{qistate}
	\chi^{A|B}=\sum_ip_i\sigma^A_i\otimes\ketbra{i}{i}^B.
	\end{equation}
	Here, $\sigma^A_i$ is an arbitrary quantum state on Alice's side and the state $\ket{i}^B$ belongs to the local incoherent basis of Bob. 
	
	\subsection{\label{appendix1:sub2}Quantification of quantum coherence in single and bipartite systems}
	Several coherence measures have been proposed, for quantifying the degree of coherence in both single systems and bipartite systems.
	
	Regarding the degree of coherence in a single system, we adopt the most popular quantifier: the \textit{relative entropy of coherence} (REC). The REC captures how far a given state $\rho$ is from the set of incoherent states,
	\begin{equation}\label{REC}
	C_r(\rho)=\min_{\chi\in\mathcal{I}}S(\rho\|\chi),
	\end{equation}
	where $S(\rho\|\chi)$ denotes the \textit{relative entropy}  between two quantum states $\rho$ and $\chi$,
	\begin{equation}
	S(\rho\|\chi)=\textmd{Tr}(\rho\log_2\rho-\rho\log_2\chi), 
	\end{equation}
	and the minimization in Eq.~(\ref{REC}) is taken over all incoherent states. Another representation of REC is
	\begin{equation}
	C_r(\rho)=S[\Delta(\rho)]-S(\rho),
	\end{equation}
	where $\Delta$ denotes the dephasing operation in the incoherent basis, and $S(\rho)$ denotes the von Neumann entropy of a quantum state $\rho$,
	\begin{equation}
	S(\rho)=-\textmd{Tr}(\rho\log_2\rho).
	\end{equation}
	The REC has operational significance as it equals the \textit{distillable coherence} (DC) $C_d$~\cite{OperationalResourceTheoryofCoherence}.
	
	Regarding the quantification of coherence on one subsystem in a bipartite system, the \textit{quantum-incoherent relative entropy of coherence} ($\mathcal{QI}$ REC) is defined as~\cite{AssistedDistillationofQuantumCoherence}
	\begin{equation}\label{QIREC}
	C^{A|B}_r(\rho^{AB})=\min_{\chi^{A|B}\in\mathcal{I}^{A|B}}S(\rho^{AB}\|\chi^{A|B}),
	\end{equation}
	where the minimum is taken over the set of $\mathcal{QI}$ states. The $\mathcal{QI}$ REC $C^{A|B}_r$ captures how close a quantum state is from the set of $\mathcal{QI}$ states. Another expression is \cite{AssistedDistillationofQuantumCoherence} 
	\begin{equation}C^{A|B}_r(\rho^{AB})=S[\Delta^B(\rho^{AB})]-S(\rho^{AB}),\end{equation}
	where $\Delta^B$ denotes dephasing in the incoherent basis of Bob.
	
	\subsection{\label{appendix1:sub3}Free operations in the resource theory of quantum coherence}
	The free operations in the QRT of coherence in a single system are operations that do not create coherence from incoherent states, 
	\begin{equation}
	\Lambda(\rho)\in\mathcal{I},\quad \forall \rho\in\mathcal{I},
	\end{equation}
	where $\mathcal{I}$ denotes the set of incoherent states. Such operations constitute the largest possible set that are free and referred to as \textit{maximally incoherent operations}. One subset of such free operations are incoherent operations, which were first introduced in~\cite{QuantifyingCoherence}, specified by a set of Kraus operators $\{K_n\}$, satisfying that each of its Kraus operators is incoherent, 
	\begin{equation}
	K_n\mathcal{I}K_n^{\dag}\subset\mathcal{I}^*,\quad \forall n, 
	\end{equation}
	where $\mathcal{I}^*$ denotes the set of diagonal semi-definite Hermitian operators. A general \textit{completely positive and trace preserving} (CPTP) map $\Lambda$ is incoherent if there exists at least one incoherent Kraus representation. Then the \textit{dephasing-covariant incoherent operations }(DIO) are maps $\Lambda$ which commute with the dephasing operation $\Delta$, i.e., $\Delta[\Lambda(\rho)]=\Lambda[\Delta(\rho)]$. And finally, the \textit{strictly incoherent operations} form the smallest set of free operations, where both $K_n$ and $K_n^\dag$ are incoherent operators.
	
	Regarding the resource theory of coherence in a bipartite scenario, where the coherence of one subsystem is viewed as resource, the \textit{local quantum-incoherent operations and classical communications} (LQICC) protocol was first introduced in \cite{AssistedDistillationofQuantumCoherence}. In a bipartite system involving Alice and Bob, Bob is restricted to perform only local incoherent operations while Alice can perform arbitrary quantum operations on her system. Classical communications between them are allowed. The $\mathcal{QI}$ REC has the operational meaning that it upper bounds the optimal generation rate of the maximally coherent state $\ket{\Phi_2}$ on Bob's side in the LQICC protocol. The \textit{distillable coherence of collaboration} (DCC) was first introduced in \cite{AssistedDistillationofQuantumCoherence},
	\begin{equation}\label{DCC}
	C^{A|B}_{d}\left(\rho^{AB}\right)= \sup\left\{R:\lim_{n\rightarrow\infty}\left(\inf_{\Lambda} \|\Lambda\left[\left(\rho^{AB}\right)^{\otimes n}\right]-\Phi_2^{\otimes\lfloor Rn\rfloor}\|\right)=0\right\},
	\end{equation}
	where the infimum is taken over all $\textmd{LQICC}$ operations $\Lambda$ and $\lfloor x\rfloor$ returns the maximum integer no larger than $x$. The DCC is upper bounded by the $\mathcal{QI}$ REC. For pure states, $C^{A|B}_d(\ket{\Phi^{AB}})=C^{A|B}_r(\ket{\Phi^{AB}})$~\cite{AssistedDistillationofQuantumCoherence}.
	
	\subsection{\label{appendix1:sub4}Relation between the $\mathcal{QI}$ REC and the SIC}
	First we introduce a relation between the $\mathcal{QI}$ REC and the \textit{steering induced coherence} (SIC).\\
	
	\textit{Proposition 1}---For a bipartite state $\rho^{AB}$, the SIC is upper bounded by the $\mathcal{QI}$ REC; i.e., we have 
	\begin{equation}
	C^{A|B}_r(\rho^{AB})\,\geq\,\bar{C}^B_r(\rho^{AB}).
	\end{equation}

	\textit{Proof}---Note that the $\mathcal{QI}$ REC can be expressed as
	\begin{equation}
	C^{A|B}_r(\rho^{AB})=\min_{\chi^{A|B}\in\mathcal{I}^{A|B}}S(\rho^{AB}||\chi^{A|B})=S[\rho^{AB}||\Delta^B(\rho^{AB})].
	\end{equation} 
	
	Recall that the quantum relative entropy has many important properties, such that \cite{vedral2002role,VedrP98}:
	\begin{subequations}\label{contractivity}
		\begin{align}
		&(a)\quad S[\Lambda(\rho)||\Lambda(\sigma)]\,\leq\, S(\rho||\sigma), \\
		&(b)\quad \sum_i p_i S\left(K_i\rho K^\dag_i/p_i||K_i\sigma K^\dag_i/q_i\right)\,\leq\,\sum_i S\left(K_i\rho K^\dag_i||K_i\sigma K^\dag_i\right), \\
		&(c)\quad S\left(\sum_iP_i\rho P_i||\sum_iP_i\sigma P_i\right)=\sum_i S\left(P_i\rho P_i||P_i\sigma P_i\right),\\
		&(d)\quad S\left(P_i\otimes\rho||P_i\otimes\sigma\right)=S(\rho||\sigma)
		\end{align}
	\end{subequations}
	where $p_i=\textmd{Tr}\left(K_i\rho K^\dag_i\right)$, $q_i=\textmd{Tr}\left(K_i\sigma K^\dag_i\right)$, and $\{P_i\}$ is a set of orthogonal projectors.
	
	Note that $\{\mathbf{M}_n^A=\ketbra{n}{n}^A\}$ is a set of projectors on Alice's system, corresponding to the measurement outcome $n$. Thus with property (a) in Eq.~(\ref{contractivity}), we have
	\begin{equation} 
	C^{A|B}_r(\rho^{AB})=S[\rho^{AB}||\Delta^B(\rho^{AB})]\,\geq\, S\left[\sum_n \mathbf{M}_n^A\rho^{AB}\mathbf{M}_n^A||\sum_n \mathbf{M}_n^A\Delta^B(\rho^{AB})\mathbf{M}_n^A\right].
	\end{equation}
	
	Following (c) in Eq.~(\ref{contractivity}), if we denote $\delta^B_n=\bra{n}\rho^{AB}\ket{n}$, $p_n=\textmd{Tr}(\delta^B_n)$, and $\rho^B_n=\frac{\delta^B_n}{p_n}$, we have
	\begin{equation}
	S\left[\sum_n \mathbf{M}_n^A\rho^{AB}\mathbf{M}_n^A||\sum_n \mathbf{M}_n^A\Delta^B(\rho^{AB})\mathbf{M}_n^A\right]=\sum_nS[\mathbf{M}_n^A\otimes\delta^B_n||\mathbf{M}_n^A\otimes\Delta^B(\delta^B_n)].
	\end{equation}
	
	Then with property (b) in Eq.~(\ref{contractivity}), we can obtain
	\begin{equation}
	\sum_nS\left[\mathbf{M}_n^A\otimes\delta^B_n||\mathbf{M}_n^A\otimes\Delta^B\left(\delta^B_n\right)\right]=\sum_nS\left[\delta^B_n||\Delta^B\left(\delta^B_n\right)\right]\,\geq\, \sum_np_nS\left[\rho^B_n||\Delta^B\left(\rho^B_n\right)\right]=\sum_np_nC_r^B\left(\rho^B_n\right).
	\end{equation}
	
	Note that $\sum_nC_r^B(\rho^B_n)$ is the average coherence that Bob can obtain with Alice's measurement choice $\{\mathbf{M}_n^A=\ketbra{n}{n}^A\}$ and classical communications. No matter what measurement Alice actually chooses, Bob will reach an average coherence no greater than $C^{A|B}_r(\rho^{AB})$. Hence, we have $C^{A|B}_r(\rho^{AB})\,\geq\,\bar{C}^B_r(\rho^{AB})$.
	
	\subsection{\label{appendix1:sub5}Difference between the $\mathcal{QI}$ REC and the quantum correlations}
	Note that the $\mathcal{QI}$ REC is essentially different from the measures of quantum correlations. First, any state that is not quantum-incoherent has nonzero $\mathcal{QI}$ REC. The difference between the measures of quantum correlations and the $\mathcal{QI}$ REC is that the latter is basis-dependent. One of the most popular measures of quantum correlations is the \textit{relative entropy of quantum discord}~\cite{ollivier2001quantum,henderson2001classical,RevModPhysdiscord}, defined as
	\begin{equation}\label{discordmeasure}
	D(\rho^{AB})=\min_{\delta^{AB}\in\mathcal{CC}}S(\rho^{AB}||\delta^{AB})
	\end{equation}
	where $\mathcal{CC}$ denotes the set of \textit{classical correlated states} that can be written in the form of the sum of projectors 
	\begin{equation}
	\delta^{AB}=\sum_{k,l}p_{kl}\ketbra{k}{k}^A\otimes\ketbra{l}{l}^B.
	\end{equation}
	Now consider a family of bipartite states 
	\begin{equation}
	f^{AB}=\sum_{k,m} p_{km} \ketbra{k}{k}^A\otimes\ketbra{m}{m}^B, 
	\end{equation}
	where $\ket{k}^A$ is any orthonormal basis of Alice, and $\ket{m}^B$ is any orthonormal basis of Bob that is not incoherent. Obviously, $f^{AB}$ is not quantum-incoherent and thus has a nonzero $\mathcal{QI}$ REC, while $f^{AB}$ has zero discord or entanglement. Moreover, we can also construct states that are quantum-incoherent but have nonzero quantum discord.
	
	\subsection{\label{appendix1:sub6}Concurrence}
	The concurrence is an entanglement monotone, defined for a mixed state of two qubits as:
	\begin{equation}\label{Concurrence}
	E^{AB}_c(\rho)\equiv \max\left(0, \lambda_{1}-\lambda_{2}-\lambda_{3}-\lambda_{4}\right),
	\end{equation}
	where $\lambda_{1},...,\lambda_{4}$ are the eigenvalues of the Hermitian matrix $R={\sqrt{{\sqrt{\rho}}{\tilde{\rho}}{\sqrt{\rho}}}}$, with
	\begin{equation}\label{sigmaxin}
	{\tilde{\rho}}=(\sigma_{y}\otimes\sigma_{y})\rho^{*}(\sigma_{y}\otimes\sigma_{y}),
	\end{equation}
	where $\rho^{*}$ denotes the spin-flipped state of $\rho$, $\sigma_y$ a Pauli spin matrix, and the eigenvalues are listed in decreasing order.
	
	\section{\label{sec:appendix2}Dynamical behaviors of the information carriers based on coherence in open systems}
	We first show that both the behaviors of the local coherence (of a single system) and the extended coherence (of the system and an ancilla) are monotonic under \textit{incoherent open system dynamics} (IOSD).
	
	A theoretical work~\cite{chanda2016delineating} showed that CP-nondivisibility of non-Markovian dynamics and monotonic behaviors of quantum coherence measures under \textit{incoherent completely positive and trace preserving} (ICPTP) maps can be used to detect and quantify the non-Markovianity of an IOSD. Obviously the REC behaves monotonically in an IOSD.
	\\
	
	\textit{Lemma} 1---The amount of REC of a quantum system decreases monotonically in an IOSD $\{\Lambda_t\}$; i.e., for any $t\,\geq\, s\,\geq\,0$, we have 
	\begin{equation}
	C_r[\Lambda_t(\rho)]\,\leq\, C_r[\Lambda_s(\rho)]. 
	\end{equation}
	
	\textit{Proof}---The family of IOSDs will preserve incoherent states; i.e., for each $\rho\in\mathcal{I}$, we have
	\begin{equation}
	\Lambda_t(\rho)\in\mathcal{I},
	\end{equation}
	for any $t\,\geq\, 0$. Let us denote the quantum states after evolution time $s$ and $t$ ($0\,\leq\, s\,\leq\, t$) are $\rho_s$ and $\rho_t$. Then the REC of $\rho_t$ and $\rho_s$ can be evaluated as
	\begin{subequations}
		\begin{align}
		&C_r(\rho_t)=S[\rho_t||\Delta(\rho_t)],\\ 
		&C_r(\rho_s)=S[\rho_s||\Delta(\rho_s)].
		\end{align}
	\end{subequations}
	Note that if $\{\Lambda_t\}$ is Markovian, we have $\Lambda_t=\Lambda_{t,s}\Lambda_{s,0}$, and $\Lambda_{t,s}$ is CPTP. Hence, we can use the relation
	\begin{equation}
	S[\rho_t||\Delta(\rho_t)]\,\leq\, S[\Lambda_{t,s}(\rho_s)||\Lambda_{t,s}\Delta(\rho_s)],
	\end{equation} 
	and the definition of coherence measure yields
	\begin{equation}
	C_r(\rho_t)=\min_{\delta_t\in\mathcal{I}}S(\rho_t||\delta_t)\,\leq\, S[\rho_t||\Lambda_{t,s}\Delta(\rho_s)].
	\end{equation}
	Then we can obtain 
	\begin{equation}\label{inequalREC}
	C_r(\rho_t)\,\leq\, S[\Lambda_{t,s}(\rho_s)||\Lambda_{t,s}\Delta(\rho_s)]\,\leq\, C_r(\rho_s).
	\end{equation}
	We have also used the property in Eq.~(\ref{contractivity}) that the quantum relative entropy is contractive under CPTP maps. Thus we complete the proof.\\
	
	When considering the overall coherence of the open system and an ancilla, we can obtain a similar conclusion. The theoretical work in~\cite{he2017non} proposed an alternative non-Markovianity measure based on the REC, which uses the whole coherence
	in an extended Hilbert space (which is referred to as extended coherence) constituted by both the open system and its ancilla. They find that the proposed measure can capture effectively the characteristics of non-Markovianity of incoherent open quantum processes, including both the phase-damping channel and the amplitude-damping channel.
	\\
	
	\textit{Lemma} 2---The amount of extended coherence of the open system (Alice) and an ancilla (Bob) decreases monotonically in an IOSD $\{\Lambda_t^A\}$ on Alice; i.e., for any $t\,\geq\, s\,\geq\,0$, we have 
	\begin{equation}
	C_r[(\Lambda_t^A\otimes \mathbb{I}^B)(\rho^{AB}_0)]\,\leq\, C_r[(\Lambda_s^A\otimes \mathbb{I}^B)(\rho^{AB}_0)].
	\end{equation}
	
	\textit{Proof}---Following \cite{he2017non} and the proof of \textit{Lemma} 1, with 
	\begin{subequations}
		\begin{align}
		&C_r(\rho_t^{AB})=S[\rho_t^{AB}||\Delta^{AB}(\rho_t^{AB})],\\ &C_r(\rho_s^{AB})=S[\rho_s^{AB}||\Delta^{AB}(\rho_s^{AB})],
		\end{align} 
	\end{subequations}
	we have 
	\begin{equation}
	C_r(\rho_t^{AB})\,\leq\, S[\Lambda_{t,s}^A(\rho_s^{AB})||\Lambda_{t,s}^A\Delta^{AB}(\rho_s^{AB})]\,\leq\, C_r(\rho_s^{AB}).
	\end{equation}
	Thus we see the dynamical behavior of the extended coherence during an IOSD.\\
	
	From the above two Lemmas, we can see that both the coherence of the open system and the extended coherence of the open system and the ancilla decrease monotonically during an IOSD, and they can be used for efficiently detecting non-Markovianity in IOSDs. However, there are dynamics that are not incoherent. Hence, both of these information carriers will not decrease monotonically in these processes. We then prove \textit{Theorem} 1, showing the monotonic behaviors of the $\mathcal{QI}$ REC during general open system dynamics.
	\\
	
	\textit{Proof of Theorem} 1---First we prove that the $\mathcal{QI}$ REC decreases monotonically during a Markovian evolution on Alice. The quantum states (initially $\rho^{AB}$) after evolution time $s$ and $t$ ($0\leq s\leq t$) are $\rho_s^{AB}$ and $\rho_t^{AB}$. We can express the $\mathcal{QI}$ REC as follows,
	\begin{equation}
	C_r^{A|B}(\rho_{t}^{AB})=S[\rho_t^{AB}\|\Delta^B(\rho_t^{AB})]=S[\Lambda^A_{t,s}(\rho_s^{AB})\|\Delta^B\Lambda^A_{t,s}(\rho_s^{AB})].
	\end{equation}
	As $\Delta^B$ acts only on $B$, and $\Lambda^A_{t, s}$ acts only on $A$, we have
	\begin{equation}
	\Delta^B\Lambda^A_{t,s}(\cdot)=\Lambda^A_{t,s}\Delta^B(\cdot).
	\end{equation}
	Combining the contractive property, if the intermediate map $\Lambda^A_{t,s}$ is CP, we have
	\begin{equation}
	S[\rho_t^{AB}\|\Delta^B(\rho_t^{AB})]\leq S[\rho_s^{AB}\|\Delta^B(\rho_s^{AB})]. 
	\end{equation}
	Thus during a Markovian process, the $\mathcal{QI}$ REC decreases monotonically.
	
	Then we prove that the SIC decreases monotonically during a Markovian evolution on Alice. It is well known that any CPTP operation on Alice can be constructed by first implementing a unitary $U$ to $A$ and an ancilla $A'$, then discarding $A'$. The corresponding mathematical formulation can be described by
	\begin{equation}\label{CPTPunitary}
	\Lambda({\rho^A})=\mathrm{Tr}_{A'}[U(\rho^A\otimes\rho^{A'})U^\dag],
	\end{equation}
	where $\rho^A$ and $\rho^{A'}$ denote the quantum states of $A$ and $A'$. 
	
	Note that if Bob shares a multipartite state $\rho^{A_1...A_n B}$ made of $n$ parties $A_1, A_n,... A_n$, the generalized steering-induced coherence can be expressed as
	\begin{equation}
	\bar{C}_r^B(\rho^{A_1...A_nB})=\max_{\mathcal{M}_{A_1...A_n}}\sum_mp_mC_r(\rho^B_m),
	\end{equation}
	where $\mathcal{M}_{A_1...A_n}$ denotes the collective projective measurements across $n$ particles. In our case, Alice ($A$), Bob and the ancilla $A'$ share a tripartite state before the unitary $U$, where $A$ and $A'$ are uncorrelated. The overall state admits the form
	\begin{equation}\label{tripartitestate}
	\rho^{A'AB}=\rho^{A'}\otimes\rho^{AB}.
	\end{equation}
	As the collective measurement on $A$ and $A'$ will reduce to a \textit{positive-operator valued measure} (POVM) on $A$, and the set of all POVMs on $A$ forms a strict larger set than the set of all local projective measurements, yielding higher average coherence that can be obtained on Bob's system in general. Thus, we obtain the following relation,
	\begin{equation}\label{generalrelationSIC}
	\bar{C}_r^B(\rho^{A'AB})=\max_{\mathcal{M}_{AA'}}\sum_mp_mC_r(\rho^B_m)\, \geq\, \bar{C}_r^B(\rho^{AB}).
	\end{equation}
	The above inequality is valid in a more general case when $A$ and $A'$ are correlated. 
	However, in the case when $A'$ and $A$ are product states, we have
	\begin{equation}
	\bar{C}_r^B(\rho^{A'AB})=\bar{C}^{B}_r(\rho^{AB}).
	\end{equation}
	First we present the proof of this statement. Considering a projective collective measurement, specified by 
	\begin{equation}
	\{\mathcal{M}_{A'A}^k= \ketbra{\psi_{k}}{\psi_{k}}\},
	\end{equation}
	which acts on $A$ and $A'$, where $\{\ket{\psi_k}\}$ forms a set of orthogonal normalized bases of the overall system, 
	\begin{equation}
	\ket{\psi_{k}}=\sum_{a,a'}\sqrt{r^k_{aa'}}\ket{a'}\otimes\ket{a}.
	\end{equation}
	Here $\{\ket{a'}\}$ and $\{\ket{a}\}$ are a set of orthonormal bases on $A'$ and $A$, respectively, and the normalization condition leads to
	\begin{equation}
	\sum_{a,a'}r^k_{aa'}=1.
	\end{equation}
	The reduced POVM on $A$ reads 
	\begin{equation}
	M_{k}=\sum_{a,a'} r^k_{aa'}\bra{a'}\rho^{A'}\ket{a'}\ketbra{a}{a},
	\end{equation}
	which corresponds to the auxiliary state $\rho^{A'}$. 
	The complete condition for POVM requires that
	\begin{equation}\label{POVMsum}
	\sum_kM_k=\mathbb{I},
	\end{equation}
	where $\mathbb{I}$ denotes the identity operator on $A$. Using the relation $\sum\ketbra{a}{a}=\mathbb{I}$, we have
	\begin{equation}\label{Coesum}
	\sum_{k,a'} r^k_{aa'}\bra{a'}\rho^{A'}\ket{a'}=1.
	\end{equation}
	Then after implementing the POVM, Bob can obtain the state from Alice's outcome $m_k$ according to each $M_k$. The state of Bob can be expressed as
	\begin{equation}
	\rho^B_k=\frac{\mathrm{Tr}_{A}(M_k\rho^{AB})}{\mathrm{Tr}_{AB}(M_k\rho^{AB})}. 
	\end{equation}
	Denoting $p_k=\mathrm{Tr}_{AB}(M_k\rho^{AB})$, the average coherence obtained by Bob after the implementation of the POVM can be expressed as 
	\begin{equation}
	\bar{C}_r^B(\rho^{A'AB}|M_{A'A})=\sum_k p_k C_r(\rho^B_k).
	\end{equation}
	If we denote the state $\rho^B_a$ as Bob's state after Alice's projective measurement $\ket{a}$, then we have
	\begin{equation}
	\rho^B_k=\frac{1}{p_k}\sum_{a,a'} r^k_{aa'}\bra{a'}\rho^{A'}\ket{a'}p_a\rho^B_a, 
	\end{equation}
	where $p_a=\mathrm{Tr}(\ketbra{a}{a}\rho^{AB})$. Then the average coherence becomes
	\begin{equation}
	\bar{C}_r^B(\rho^{A'AB}|M_{A'A})=\sum_kp_k C_r\left(\frac{1}{p_k}\sum_{a,a'} r^k_{aa'}\bra{a'}\rho^{A'}\ket{a'}p_a\rho^B_a\right).
	\end{equation}
	As the convexity of REC~\cite{QuantifyingCoherence},
	\begin{equation}
	C_r\left(\sum_i p_i\rho_i\right)\,\leq\,\sum_ip_i C_r(\rho_i),
	\end{equation}
	the REC will not increase under mixture of quantum states,
	\begin{equation} 
	\sum_kp_k C_r\left(\frac{1}{p_k}\sum_{a,a'} r^k_{aa'}\bra{a'}\rho^{A'}\ket{a'}p_a\rho^B_a\right)\,\leq\,\sum_{a,a',k}r^k_{aa'}\bra{a'}\rho^{A'}\ket{a'}p_a C_r(\rho^B_a).
	\end{equation}
	Combining Eq.~(\ref{POVMsum}) and Eq.~(\ref{Coesum}), we have 
	\begin{equation}\label{inequalSICF}
	\sum_{a,a',k}r^k_{aa'}\bra{a'}\rho^{A'}\ket{a'}p_a C_r(\rho^B_a)=\sum_ap_aC_r(\rho^B_a),
	\end{equation}
	the right hand of inequality~(\ref{inequalSICF}) is the average coherence, obtained by the projective measurement $\{\ketbra{a}{a}\}$ on $A$. From the inequality we can see that the average coherence, obtained from any collective projective measurement on $A'$ and $A$, is no greater than the average coherence, which is obtained from a proper projective measurement on $A$ if $A$ and $A'$ are product states. After the CPTP operations on $A$, which is realized by the unitary on $A'$ and $A$, the maximum of average coherence obtained from the collective projective measurement remains unchanged, i.e.,
	\begin{equation}
	\bar{C}_r^B(\rho^{A'AB})=\bar{C}_r^B(\rho^{A'AB}_{U}),
	\end{equation}
	where $\rho^{A'AB}_{U}=U^{AA'}\rho^{A'AB}(U^{AA'})^\dag$. The average coherence obtained from the collective projective measurement on $A$ and $A'$ is upper bounded by $\bar{C}_r^B(\rho^{AB})$. After the unitary $U$, we have
	\begin{equation}
	\bar{C}_r^B(\rho^{A'AB}_{U})=\bar{C}_r^B(\rho^{AB}).
	\end{equation}
	When tracing over $A'$, the set of projective measurements on $A$ is a strict subset of the reduced POVMs corresponding to collective projective measurements on $A$ and $A'$. Thus, we have 
	\begin{equation}
	\bar{C}_r^B(\rho^{AB}_{U})\,\leq\,\bar{C}_r^B(\rho^{A'AB}_{U}),
	\end{equation}
	where $\rho^{AB}_{U}$ denotes the final states of Alice and Bob after the implementation of $U$,
	\begin{equation}
	\rho^{AB}_{U}=\mathrm{Tr}_{A'}(\rho^{A'AB}_{U}).
	\end{equation}
	Using the fact that any CPTP map on A can be constructed by unitary interaction between $A$ and an uncorrelated $A'$, as denoted in Eq.~(\ref{CPTPunitary}), the SIC of $B$ will decrease under a CPTP map on $A$,
	\begin{equation}
	\bar{C}_r^B[\Lambda^A(\rho^{AB})]\,\leq\,\bar{C}_r^B(\rho^{AB}).
	\end{equation}
	Thus we complete the proof. This result shows that the Markovianity on Alice's evolution will reduce the steerability of Alice to Bob's state, shrinking the accessible states of Bob, while the local state of Bob $\rho^B$ will remain unchanged.
	
	\section{\label{sec:appendix3}Numerical simulations for different processes}
	In this section we present the numerical simulations to show the behaviors of the local coherence of a single system, the extended coherence with an ancilla, the $\mathcal{QI}$ REC, and the SIC, under different non-Markovian quantum dynamics. To simulate the behavior of the REC of a single qubit system, we use the initial pure state 
	\begin{equation}
	\ket{\psi_0}^A=\frac{1}{2}\left(\sqrt{3}\ket{0}+\ket{1}\right). 
	\end{equation}
	And to simulate the dynamical behaviors of the extended coherence with an ancilla, the $\mathcal{QI}$ REC, and the SIC on a bipartite system, we choose the initial two-qubit entangled state
	\begin{equation}
	\ket{\psi_0}^{AB}=\frac{1}{4}\left(\sqrt{6}\ket{00}+\sqrt{2}\ket{01}+\sqrt{2}\ket{10}+\sqrt{6}\ket{11}\right).
	\end{equation}
	We consider two kinds of quantum dynamics: the amplitude-damping channels and the multiple decoherence channels.

	\subsection{\label{appendix3:sub1}Amplitude-damping channels}
	We now consider the single-qubit amplitude-damping channels modeled by the Hamiltonian
	\begin{equation}\label{amhamiltonian}
	H_{\mathrm{tot}}=\frac{1}{2}\omega_0\sigma_z+\sum_i\omega_ia^\dag_ia_i+\sum_i\left(g_i\sigma_+a_i+g_i^*\sigma_-a_i^\dag\right)
	\end{equation}
	where, $\sigma_+$ and $\sigma_-$ are the raising and lowering operators for the qubit. The master equation corresponding to the Hamiltonian in Eq.~(\ref{amhamiltonian}) is given by
	\begin{equation}\label{ammaster}
	\frac{d}{dt}\rho_t=-\frac{\mathrm{i}}{4}S(t)\left[\sigma_z,\rho\right]+\gamma(t)\left(\sigma_-\rho_t\sigma_+-\frac{1}{2}\{\sigma_+\rho_t\sigma_--\rho_t\}\right)
	\end{equation}
	where the qualities $S(t)$ and $\gamma(t)$ read
	\begin{subequations}
		\begin{align}
		&S(t)=-2\mathrm{Im}\frac{\dot{G}(t)}{G(t)},\\
		&\gamma(t)=-2\mathrm{Re}\frac{\dot{G}(t)}{G(t)},
		\end{align}
	\end{subequations}
	and the decoherence function $G(t)$ depends on the spectral density $J(\omega)$. Considering a Lorentzian shape spectral density,
	\begin{equation}
	J(\omega)=\frac{\gamma_0\lambda^2}{[(\omega_0+\delta-\omega)^2+\lambda^2]},
	\end{equation}
	and letting $\delta=0$, one obtains the decoherence function $G(t)$ as
	\begin{equation}\label{amdecoherencefunction}
	G(t)=\exp\left(\frac{-\lambda t}{2}\right)\left[\cosh\left(\frac{dt}{2}\right)+\frac{\lambda}{d}\sinh\left(\frac{dt}{2}\right)\right]
	\end{equation}
	where $d=\sqrt{\lambda^2-2\gamma_0\lambda}$. 
	
	Here, we consider both Markovian dynamics (when $\gamma_0<\lambda/2$), and non-Markovian dynamics (when $\gamma_0>\lambda/2$). The corresponding results are shown in Fig.~\ref{fig:dataA1} and Fig.~\ref{fig:dataA12}, respectively. 
	
	In the Markovian regime, we choose $\gamma_0=0.2\lambda$, the dynamical behaviors of the $\mathcal{QI}$ REC, the extended coherence, and the local coherence, with respect to different bases (as shown in Fig.~\ref{fig:dataA1}). We simulate the dynamical behaviors of the above coherence measures in different reference bases. In the Markovian regime of the amplitude-damping channels, the $\mathcal{QI}$ REC and the extended coherence behaves monotonically. However, the local coherence  behaves differently in different reference bases; i.e., the monotonicity depends on the choice of reference basis. Note that although the extended coherence of Alice and Bob behaves monotonically in all reference bases in the Markovian regime of the amplitude-damping channel, it cannot be used for detecting non-Markovianity in the general evolution as the case we have experimentally shown in the main text.
	
	In the non-Markovian regime, we choose $\gamma_0=25\lambda$, resulting in the non-Markovianity of the open system dynamics. In this case, we simulate both the $\mathcal{QI}$ REC and the SIC with respect to different bases (as shown in Fig.~\ref{fig:dataA12}). From the simulation, we can see that during the amplitude-damping channel, the non-Markovianity can be detected with both the $\mathcal{QI}$ REC and the SIC (note that in the above case the values of the SIC are coincident in all reference bases) independent of the reference basis we choose.
	
	\begin{figure}[htp]
		\label{fig:dataA1}
		\includegraphics[scale=0.18]{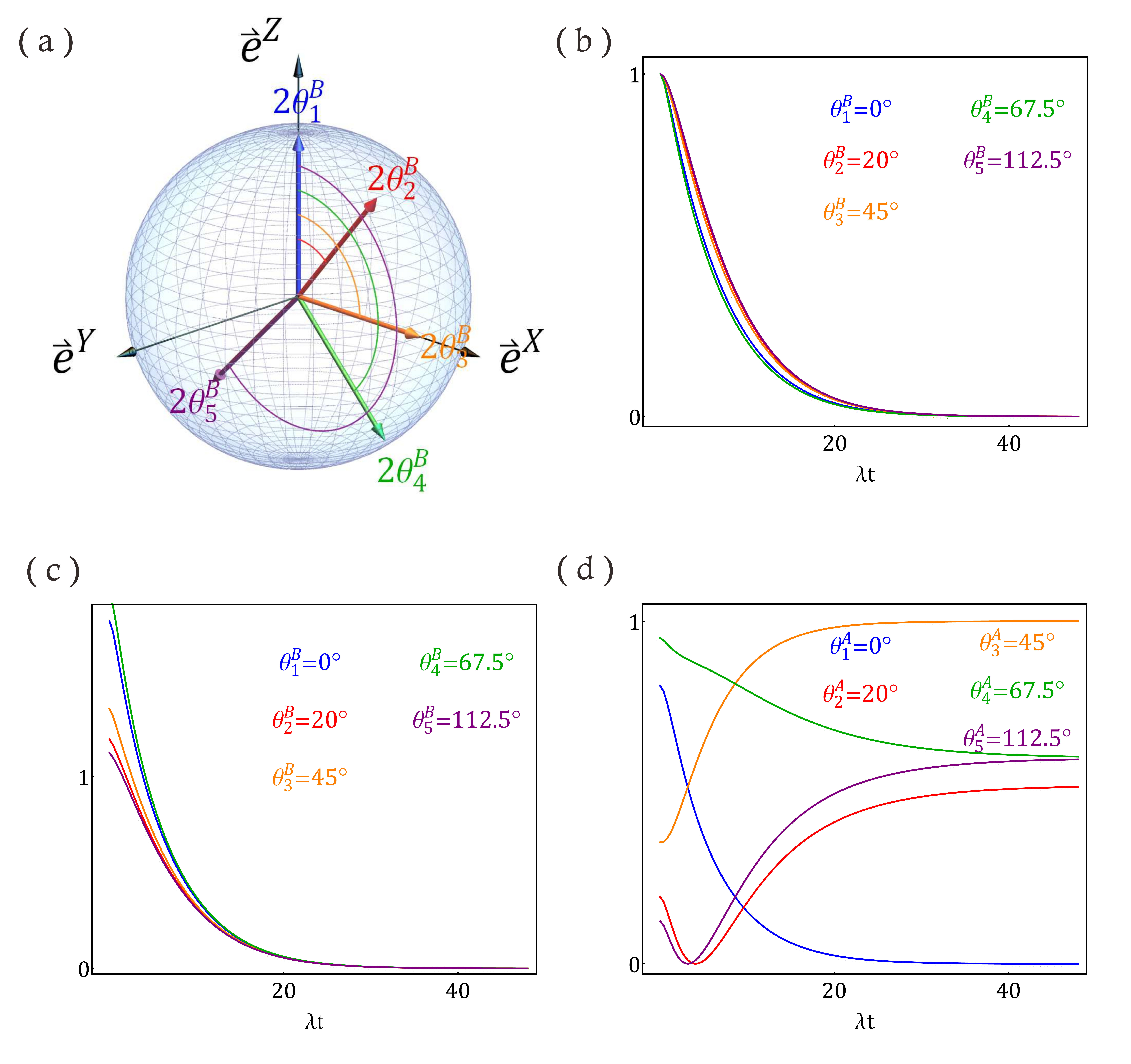}
		\caption{\label{fig:dataA1} \textbf{Theoretical simulations for the Markovian amplitude-damping channels.} In the Markovian regime, the simulations of the dynamical behaviors of the $\mathcal{QI}$ REC (b), the extended coherence (c), and the local coherence (d) are shown. The reference bases of Alice (Bob) are chosen as the eigenbasis of $\vec{\sigma}\cdot\vec{n}[\theta^{A(B)}_i]$ with different $\theta^{A(B)}_i$, where $\vec{n}(\theta_i^B)=\sin2\theta_i^B\vec{e}^X+\cos2\theta_i^B\vec{e}^Z$. During the Markovian evolution, when $\gamma_0=0.2\lambda$, both the $\mathcal{QI}$ REC and the extended coherence of $\rho^{AB}$ decrease monotonically. The dynamical behavior of the local coherence of $A$ is non-monotonic and depends on the basis we choose.}
	\end{figure}
	
	\begin{figure}[htp]
		\label{fig:dataA12}
		\includegraphics[scale=0.18]{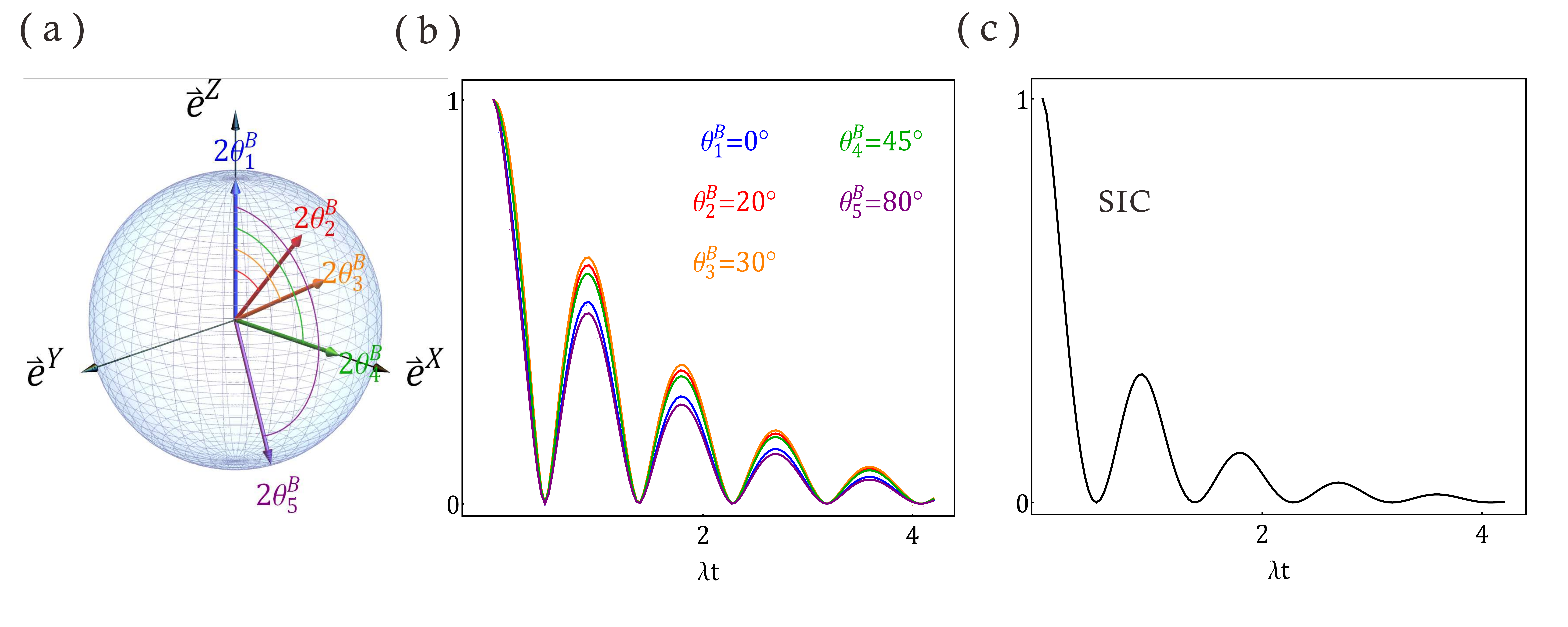}
		\caption{\label{fig:dataA12} \textbf{Theoretical simulations for the non-Markovian amplitude-damping channels.}  In the non-Markovian regime, $\gamma_0=25\lambda$, the non-Markovianity can be captured by the $\mathcal{QI}$ REC (b) and the SIC (c), where the SIC behaves exactly the same in all reference bases, chosen as the eigenbasis of $\vec{\sigma}\cdot\vec{n}(\theta^{B}_i)$ with different $\theta^{B}_i$, where $\vec{n}(\theta_i^B)=\sin2\theta_i^B\vec{e}^X+\cos2\theta_i^B\vec{e}^Z$.}
	\end{figure}

	\subsection{\label{appendix3:sub2}Multiple decoherence channels}
	The dynamics of a single qubit in multiple decoherence channels can be considered for a two-level system with the master equation
	\begin{equation}\label{randomunitarymaster}
	\frac{d}{dt}\rho_t=\frac{1}{2}\sum_{i=1}^{3}\gamma_i(t)(\sigma_i\rho_t\sigma_i-\rho_t),
	\end{equation}
	where $\sigma_i$ denotes the $i$th Pauli matrix. The dynamical map corresponding to Eq.~(\ref{randomunitarymaster}) can be exactly worked out and is given by the random unitary dynamics
	\begin{equation}\label{randomunitarymap}
	\Lambda_t(\rho)=\sum_{i=0}^3p_i(t)\sigma_i\rho\sigma_i.
	\end{equation}
	Here, for the Markovian dynamics, we set the paramaters $\gamma_i(t)$ as follows
	\begin{equation}
	\gamma_1(t)=\gamma_2(t)=\gamma_3(t)=\frac{c}{2},
	\end{equation}
	and for the non-Markovian dynamics, we set the aforementioned parameters as
	\begin{subequations}
		\begin{align}
		&\gamma_1(t)=\gamma_2(t)=\frac{c}{2},\\
		&\gamma_3(t)=\frac{c\lambda\cos ct}{2},
		\end{align} 
	\end{subequations}
	where $c>0$, and $\lambda>0$, controlling the degree of non-Markovianity.
	
	In the Markovian regime, the dynamics of the system can be exactly solved as 
	\begin{subequations}
		\begin{align}
		&p_0(t)=\frac{1+3\exp(-2ct)}{4},\\
		&p_1(t)=p_2(t)=p_3(t)=\frac{1-\exp(-2ct)}{4}.
		\end{align}
	\end{subequations}
	The numerical simulations of the dynamical behaviors of the $\mathcal{QI}$ REC, the extended coherence, and the local coherence, with respect to different bases (as shown in Fig.~\ref{fig:dataA2}). In this case, all the above coherence measures behave monotonically during the Markovian dynamics.
	
	In the non-Markovian regime, the dynamics of the system can be solved as
	\begin{subequations}
		\begin{align}
		&p_0(t)=\frac{1+\exp(-2ct)+2\exp(-ct-\lambda\sin ct)}{4},\\ 
		&p_1(t)=p_2(t)=\frac{1-\exp(-2ct)}{4}, \\
		&p_3(t)=\frac{1+\exp(-2ct)-2\exp(-ct-\lambda\sin ct)}{4}.
		\end{align}
	\end{subequations}
	We set $\lambda=3.8$. The dynamical behaviors of the $\mathcal{QI}$ REC, and the SIC are simulated with respect to different bases, as shown in Fig.~\ref{fig:dataA22}. We can see that in all reference bases chosen, the non-Markovianity can be captured by both the temporal increase of the $\mathcal{QI}$ REC and the SIC.

	\begin{figure}[htp]
		\label{fig:dataA2}
		\includegraphics[scale=0.18]{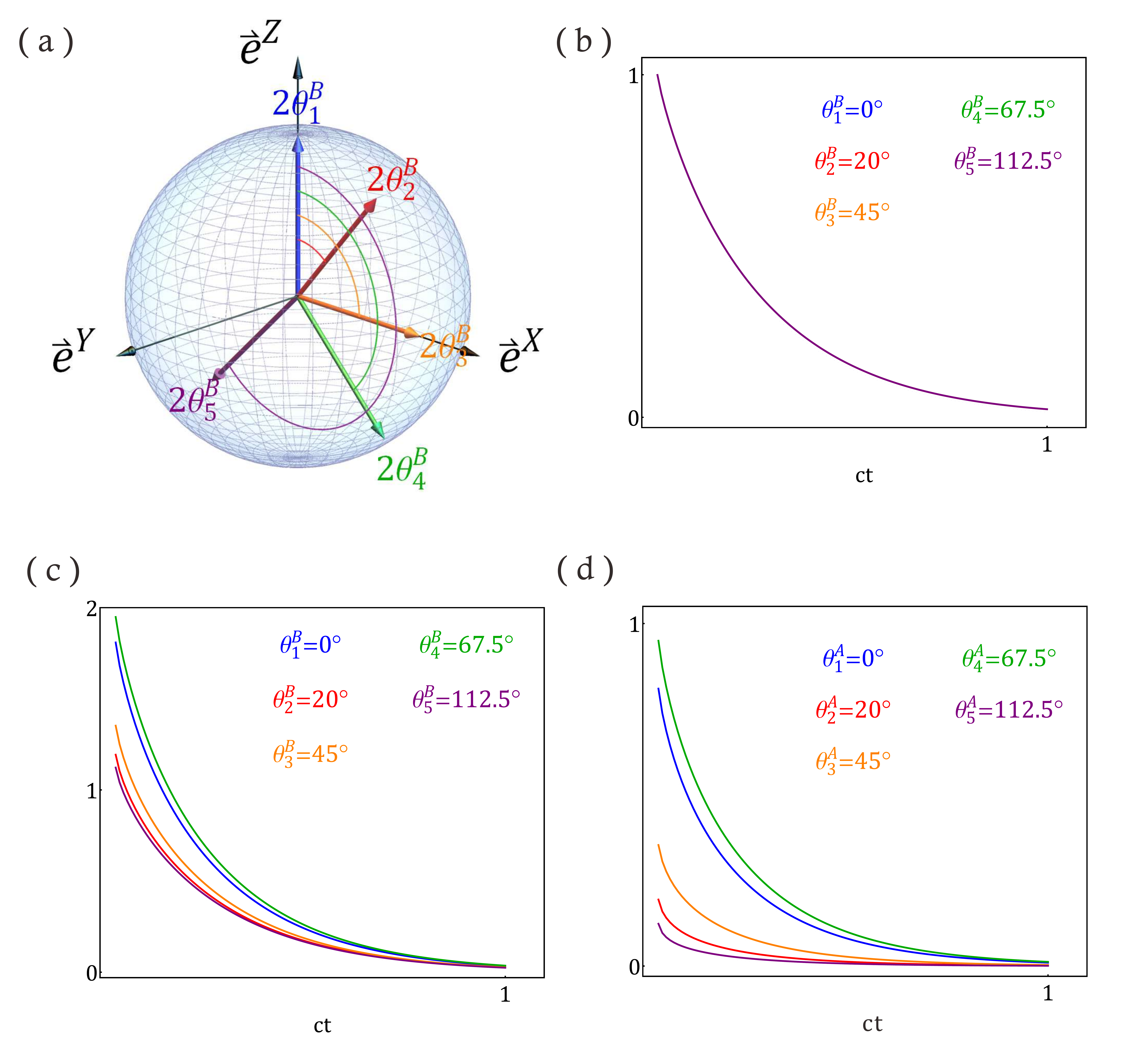}
		\caption{\label{fig:dataA2} \textbf{Theoretical simulations for the Markovian multiple decoherence channels.} In the Markovian regime, the numerical simulations of the dynamical behaviors of the $\mathcal{QI}$ REC (b), the extended coherence (c), and the local coherence (d) are shown. The reference basis of Alice (Bob) is chosen as the eigenbasis of $\vec{\sigma}\cdot\vec{n}[\theta^{A(B)}_i]$ with different $\theta^{A(B)}_i$, where $\vec{n}(\theta_i^B)=\sin2\theta_i^B\vec{e}^X+\cos2\theta_i^B\vec{e}^Z$. In the Markovian regime, all information quantifiers behave monotonically independent of the basis chosen.}
	\end{figure}
	
	\begin{figure}[htp]
		\label{fig:dataA22}
		\includegraphics[scale=0.18]{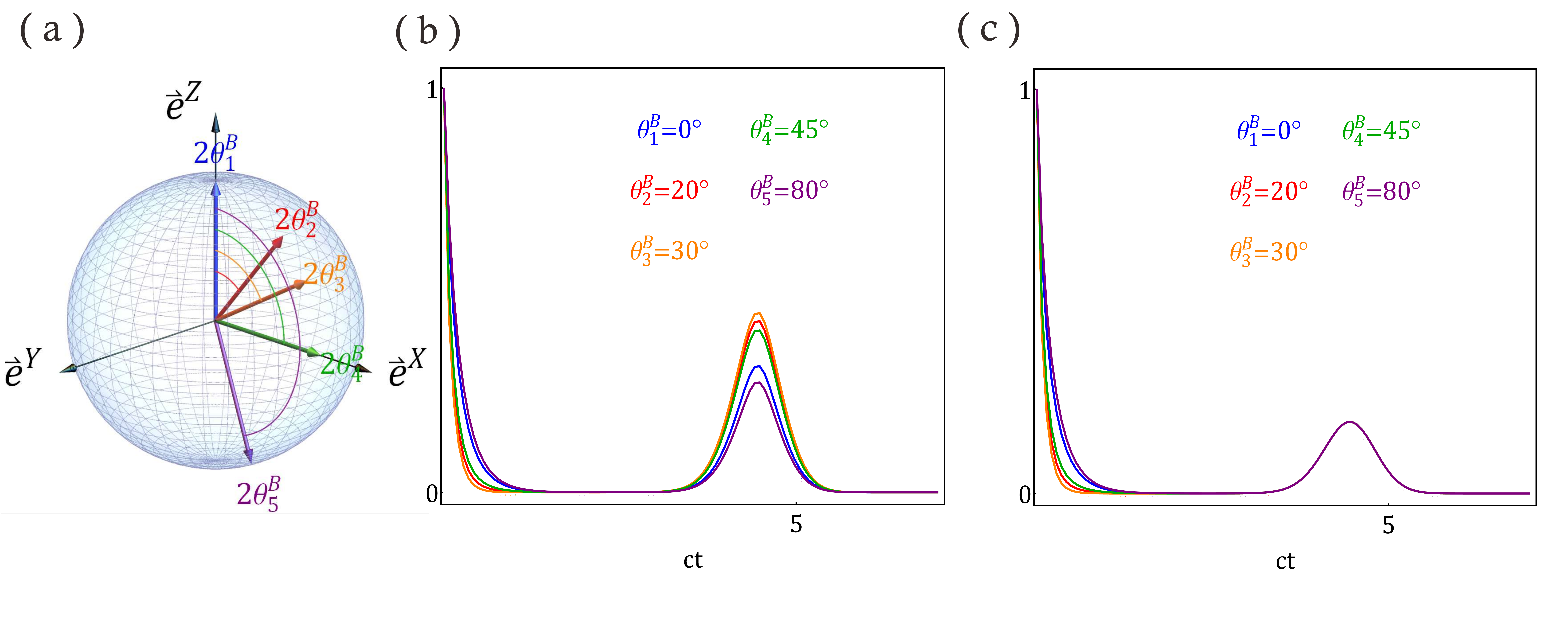}
		\caption{\label{fig:dataA22} \textbf{Theoretical simulations for the non-Markovian multiple decoherence channels.} In the non-Markovian regime, the non-Markovianity can be simultaneously captured by the temporal increase of the $\mathcal{QI}$ REC (b) and the SIC (c), in all reference bases of Bob, chosen as the eigenbasis of $\vec{\sigma}\cdot\vec{n}(\theta^{B}_i)$ with different $\theta^{B}_i$, where $\vec{n}(\theta_i^B)=\sin2\theta_i^B\vec{e}^X+\cos2\theta_i^B\vec{e}^Z$.}
	\end{figure}
	
	\section{\label{sec:appendix4}Non-Markovianity measure based on the $\mathcal{QI}$ REC}
	In this section, we define a new method for non-Markovianity measure based on the $\mathcal{QI}$ REC, which is
	\begin{equation}\label{nonMar}
	\mathcal{N}_{\mathcal{QI}}(\Lambda)=\max_{\ket{i}^B}\int_{\sigma>0}\sigma(t,\ket{i}^B),
	\end{equation}
	where
	\begin{equation}\label{sigma}
	\sigma(t,\ket{i}^B)=\frac{\partial C^{A|B_i}_r\left[\Lambda\otimes\mathbb{I}(\ketbra{\Phi}{\Phi})\right]}{\partial t},
	\end{equation}
	$C^{A|B_i}_r$ denotes the $\mathcal{QI}$ REC with respect to the reference basis $\{\ket{i}^B\}$ of Bob, and $\ket{\Phi}$ can be any pure bipartite maximally entangled state. Thus, this definition only needs optimization over all local bases of Bob's system.
	
	In order to figure out the property of this non-Markovianity measure, let us first recall two popular non-Markovian measures.
	One was defined by Breuer, Laine, and Piilo (BLP)~\cite{MeasureBreuer2009}. A dynamical map $\{\Lambda_t\}$ is Markovian if the distinguishability of any two evolving quantum states $\rho$ and $\tau$ decreases, and the associated measure for non-Markovianity measure is then defined as
	\begin{equation}\label{BLP}
	\mathcal{N}_{\mathrm{BLP}}=\max_{\rho,\tau}\int_{(\partial \|\rho_t-\tau_t\|/\partial t)>0} \frac{\partial \|\rho_t-\tau_t\|}{\partial t}dt.
	\end{equation}
	Here $\rho$ and $\tau$ denote the initial pairs of quantum states, and $\|\cdot\|$ denotes the trace distance. However, this involves a formidable optimization over all pairs of density operators, which is relatively harder to carry out when working with a high-dimensional system.
	
	The other one was proposed by Rivas, Huelga, and Plenio (RHP) in~\cite{EntanglementRivas2010}:
	\begin{equation}\label{RHP}
	\mathcal{N}_{\mathrm{RHP}}=\int_{0}^{\infty} \lim_{\varepsilon\rightarrow0} \frac{\mathrm{Tr}\|\Lambda_{t+\varepsilon,t}\otimes\mathbb{I}(\ketbra{\Phi}{\Phi})\|-1}{\varepsilon}dt
	\end{equation}
	where $\ket{\Phi}$ denotes a maximally entangled state shared by the open system and ancilla. However, this approach needs the computation of the transition map $\Lambda_{t+\varepsilon,t}$, which cannot be evaluated in general. Moreover, an entanglement measure is often difficult to evaluate itself especially in a high-dimensional system.

	\section{\label{sec:appendix5}Experimental aspects}
	\subsection{\label{appendix5:sub1}State preparation}
	In the state preparation module (\uppercase\expandafter{\romannumeral1}), two type-I phase-matched $\beta$-barium borate (BBO) crystals, whose optical axes are normal to each other, are pumped by a continuous-wave $\textmd{Ar}^+$ laser at 351.1 nm, with a power of around 50 mW, for the generation of photon pairs with a central wavelength at $\lambda$=702.2 nm via a \textit{spontaneous parametric down-conversion process} (SPDC). A half-wave plate working at 351.1 nm set before the lense and BBO crystals is used to control the polarization of the pump laser. The two polarization-entangled photons are then separately distributed through two single-mode fibers (SMF), where one represents Bob and the other Alice. Two interference filters with a 4 nm \textit{full width at half maximum }(FWHM) are placed to filter out proper transmission peaks. HWPs at both ends of the SMFs are used to control the polarization of both photons. A quarter-wave plate in Bob's arm is used to compensate the phase for the desired prepared state. A $\textmd{Fabry-P}\acute{\textmd{e}}\textmd{rot}$ cavity which is 0.06 mm thick and coated with a partial reflecting coating on each side at 702.2 nm (actually the experimental accessible FP cavity is coated with a reflectivity of around 0.85 of both sides at 780 nm, which is close to the value of the reflectivity at 702.2 nm) can be inserted into Alice's arm to change her initial environment. The setup can generate arbitrary pure bipartite states
	\begin{equation}\label{purestate}
	\ket{\Psi^{AB}}=\ket{\psi(\theta)}^{AB}\otimes\ket{\chi}^A,
	\end{equation}
	where $\ket{\psi(\theta)}^{AB}$ denotes the entangled pure states shared by Alice and Bob,
	\begin{equation}
	\ket{\psi(\theta)}^{AB}=\cos2\theta\ket{00}+\sin2\theta\ket{11},
	\end{equation}
	with arbitrary tunable $\theta$, and $0\equiv H$, $1\equiv V$, representing an incoherent basis. The maximally entangled state $\ket{\psi(\frac{\pi}{8})}$ can be prepared with a fidelity of $0.985$, with an interference visibility $C_{DD}:C_{DA}\gtrsim 100$, where $C_{DD}\,(C_{DA})$ denotes coincident events when Alice is in the state $\ket{D}=\frac{1}{\sqrt{2}}(\ket{0}+\ket{1})$ and Bob is in the state $\ket{D}$ [ $\ket{A}=\frac{1}{\sqrt{2}}(\ket{0}-\ket{1})$]. The environmental state can be expressed as
	\begin{equation} 
	\ket{\chi}^A=\int\textbf{d}\omega f(\omega)\ket{\omega},
	\end{equation}
	which involves the amplitude $f(\omega)$ for Alice's photon in a mode with frequency $\omega$ \cite{liu2011experimental}.

	\subsection{\label{appendix5:sub2}Evolution}
	In the evolution module (\uppercase\expandafter{\romannumeral2}), all plates (QPs, QWPs and HWPs) are mounted on rotation frames that allow us to construct a dephasing process in an arbitrary orthogonal basis,
	\begin{subequations}
		\begin{align}
		&\ket{n_+(\alpha)}=\cos\alpha\ket{0}+\sin\alpha\ket{1},\\
		&\ket{n_-(\alpha)}= -\sin\alpha\ket{0}+\cos\alpha\ket{1},
		\end{align}
	\end{subequations}
	where $\alpha$ depends on the angle of the optical axis of the QPs. A QWP (rotation angle set to $\alpha$) in Alice's arm is used for phase compensation between the $\ket{n_+(\alpha)}$ and $\ket{n_-(\alpha)}$ polarized photons.  The experimental evolution admits a simple theoretical analysis which is described by a unitary transformation
	
	\begin{equation}\label{unitaryt}
	\ket{n_{\pm}(\alpha)}\otimes\ket{\omega}\xrightarrow{U(\alpha)} \exp(-\mathrm{i}n_{\pm}\omega t) \ket{n_{\pm}(\alpha)}\otimes\ket{\omega};
	\end{equation}
	the corresponding dynamical map $\Lambda_t$ takes the form,
	\begin{subequations}
		\begin{align}
		&\ketbra{n_+(\alpha)}{n_+(\alpha)}\xrightarrow{\Lambda_t}\ketbra{n_+(\alpha)}{n_+(\alpha)},\\
		&\ketbra{n_-(\alpha)}{n_-(\alpha)}\xrightarrow{\Lambda_t}\ketbra{n_-(\alpha)}{n_-(\alpha)},\\
		&\ketbra{n_+(\alpha)}{n_-(\alpha)}\xrightarrow{\Lambda_t} \kappa(t)\ketbra{n_+(\alpha)}{n_-(\alpha)},\\
		&\ketbra{n_-(\alpha)}{n_+(\alpha)}\xrightarrow{\Lambda_t} \kappa^*(t)\ketbra{n_-(\alpha)}{n_+(\alpha)},
		\end{align}
	\end{subequations}
	where the decoherence factor reads
	\begin{equation}
	\kappa(t)=\int\textbf{d}\omega |f(\omega)|^2\exp(-\mathrm{i}\Delta n\omega t),
	\end{equation}
	and $\Delta n=n_+-n_-$ denotes the nonzero difference in the refraction indices of the $\ket{n_+(\alpha)}$ and $\ket{n_-(\alpha)}$ polarized photons.
	
	All theoretical simulations are performed considering the experimental imperfections, including the experimentally prepared quantum states.
	For simulating the two aforementioned processes, $\Lambda_t^{M}$ and $\Lambda_t^{NM}$, for Markovian and non-Markovian dynamics, we have made the assumption that in the experiments of the Markovian process, the frequency distribution can be well described with a Gaussian profile with a standard deviation of $6.50\times10^{12}$ Hz (its corresponding FWHM is 3.4 nm). While the non-Markovian process can be well modeled by a sum of two Gaussians centered at two different frequencies, corresponding to wavelengths 700.6 nm and 703.3 nm with amplitudes 0.65 and 0.35. 
	
	The essential difficulty in the experiments is the phase compensation for conducting the correct evolution. In the first part of the experiments, the Markovian evolution is constructed as pure dephasing in the eigenbasis $\{\ket{n_+(20^\circ)}, \ket{n_-(20^\circ)}\}$ of $\vec{\sigma}\cdot\vec{n}_0$, where $\vec{n}_0=\cos40^{\circ}\vec{e}^X+\sin40^{\circ}\vec{e}^Z$. Hence, we rotate all QPs to $20^\circ$. In ideal case, we assume that no additional phase is introduced between $\ket{n_+(20^\circ)}$ and $\ket{n_-(20^\circ)}$. However, in our experiments, an additional phase $\phi(t)$ will be introduced and the evolution of the extended coherence and the local coherence will behave differently depending on the additional phase $\phi(t)$. For solving this problem, we insert a QWP with rotation angle $20^\circ$ to compensate the phase, removing $\phi(t)$. As we take experimental data using QPs with different lengths for each evolution time $t$, the phase compensation is performed each time when we change the lengths of the QPs. In the experiments with the non-Markovian process, since the dynamical behavior of neither the local coherence nor the extended coherence is taken into consideration, the additional phase will not play an important role in the experimental errors. Thus, the $\mathcal{QI}$ REC is more robust to phase errors in our protocols.
	
	\subsection{\label{appendix5:sub3}State tomography and coherence detection}
	In the detection module (\uppercase\expandafter{\romannumeral3}), the extinction ratio of the reflected arm of a PBS is lower than the transmissive arm. For improving the extinction ratio, we use a HWP with rotation angle set to $45^\circ$ and another PBS placed in the reflected arm, resulting in an increase in the extinction ratio. Thus the precision of the tomography process can be improved. 
	
	We use multi-mode fibers for directing photons from the free space to the detectors. The use of multi-mode fibers can increase and stabilize the collection efficiency of the photons. The power of the 351.1 nm continuous laser is set to about 50 mW, and the coincidence window is set at 4 ns, resulting in around 1000 coincident events in one second. 
	
	The overall quantum state can be reconstructed via the combination of four wave plates (two HWPs and two QWPs) and two PBSs, performing a standard two-qubit state tomography. The state of a single system can also be analyzed via two wave plates and one PBS on Alice's side, while Bob's photons are used as the trigger. Then the coherence-related measures can be calculated directly from the experimentally reconstructed quantum states $\tilde{\rho}$.

\end{widetext}


\begin{thebibliography}{68}%
	\makeatletter
	\providecommand \@ifxundefined [1]{%
		\@ifx{#1\undefined}
	}%
	\providecommand \@ifnum [1]{%
		\ifnum #1\expandafter \@firstoftwo
		\else \expandafter \@secondoftwo
		\fi
	}%
	\providecommand \@ifx [1]{%
		\ifx #1\expandafter \@firstoftwo
		\else \expandafter \@secondoftwo
		\fi
	}%
	\providecommand \natexlab [1]{#1}%
	\providecommand \enquote  [1]{``#1''}%
	\providecommand \bibnamefont  [1]{#1}%
	\providecommand \bibfnamefont [1]{#1}%
	\providecommand \citenamefont [1]{#1}%
	\providecommand \href@noop [0]{\@secondoftwo}%
	\providecommand \href [0]{\begingroup \@sanitize@url \@href}%
	\providecommand \@href[1]{\@@startlink{#1}\@@href}%
	\providecommand \@@href[1]{\endgroup#1\@@endlink}%
	\providecommand \@sanitize@url [0]{\catcode `\\12\catcode `\$12\catcode
		`\&12\catcode `\#12\catcode `\^12\catcode `\_12\catcode `\%12\relax}%
	\providecommand \@@startlink[1]{}%
	\providecommand \@@endlink[0]{}%
	\providecommand \url  [0]{\begingroup\@sanitize@url \@url }%
	\providecommand \@url [1]{\endgroup\@href {#1}{\urlprefix }}%
	\providecommand \urlprefix  [0]{URL }%
	\providecommand \Eprint [0]{\href }%
	\providecommand \doibase [0]{http://dx.doi.org/}%
	\providecommand \selectlanguage [0]{\@gobble}%
	\providecommand \bibinfo  [0]{\@secondoftwo}%
	\providecommand \bibfield  [0]{\@secondoftwo}%
	\providecommand \translation [1]{[#1]}%
	\providecommand \BibitemOpen [0]{}%
	\providecommand \bibitemStop [0]{}%
	\providecommand \bibitemNoStop [0]{.\EOS\space}%
	\providecommand \EOS [0]{\spacefactor3000\relax}%
	\providecommand \BibitemShut  [1]{\csname bibitem#1\endcsname}%
	\let\auto@bib@innerbib\@empty
	%</preamble>
	\bibitem [{\citenamefont {Brand\~ao}\ and\ \citenamefont
		{Gour}(2015)}]{ReversibleFrameworkforQuantumResourceTheories}%
	\BibitemOpen
	\bibfield  {author} {\bibinfo {author} {\bibfnamefont {F.~G. S.~L.}\
			\bibnamefont {Brand\~ao}}\ and\ \bibinfo {author} {\bibfnamefont
			{G.}~\bibnamefont {Gour}},\ }\href {\doibase 10.1103/PhysRevLett.115.070503}
	{\bibfield  {journal} {\bibinfo  {journal} {Phys. Rev. Lett.}\ }\textbf
		{\bibinfo {volume} {115}},\ \bibinfo {pages} {070503} (\bibinfo {year}
		{2015})}\BibitemShut {NoStop}%
	\bibitem [{\citenamefont {Gour}\ \emph {et~al.}(2009)\citenamefont {Gour},
		\citenamefont {Marvian},\ and\ \citenamefont {Spekkens}}]{PhysRevAMarvian}%
	\BibitemOpen
	\bibfield  {author} {\bibinfo {author} {\bibfnamefont {G.}~\bibnamefont
			{Gour}}, \bibinfo {author} {\bibfnamefont {I.}~\bibnamefont {Marvian}}, \
		and\ \bibinfo {author} {\bibfnamefont {R.~W.}\ \bibnamefont {Spekkens}},\
	}\href {\doibase 10.1103/PhysRevA.80.012307} {\bibfield  {journal} {\bibinfo
			{journal} {Phys. Rev. A}\ }\textbf {\bibinfo {volume} {80}},\ \bibinfo
		{pages} {012307} (\bibinfo {year} {2009})}\BibitemShut {NoStop}%
	\bibitem [{\citenamefont {Marvian}\ and\ \citenamefont
		{Spekkens}(2014)}]{marvian2014extending}%
	\BibitemOpen
	\bibfield  {author} {\bibinfo {author} {\bibfnamefont {I.}~\bibnamefont
			{Marvian}}\ and\ \bibinfo {author} {\bibfnamefont {R.~W.}\ \bibnamefont
			{Spekkens}},\ }\href@noop {} {\bibfield  {journal} {\bibinfo  {journal} {Nat.
				Commun.}\ }\textbf {\bibinfo {volume} {5}},\ \bibinfo {pages} {3821}
		(\bibinfo {year} {2014})}\BibitemShut {NoStop}%
	\bibitem [{\citenamefont {Brand\~ao}\ \emph {et~al.}(2013)\citenamefont
		{Brand\~ao}, \citenamefont {Horodecki}, \citenamefont {Oppenheim},
		\citenamefont {Renes},\ and\ \citenamefont
		{Spekkens}}]{PhysRevLettFernandoThermal}%
	\BibitemOpen
	\bibfield  {author} {\bibinfo {author} {\bibfnamefont {F.~G. S.~L.}\
			\bibnamefont {Brand\~ao}}, \bibinfo {author} {\bibfnamefont {M.}~\bibnamefont
			{Horodecki}}, \bibinfo {author} {\bibfnamefont {J.}~\bibnamefont
			{Oppenheim}}, \bibinfo {author} {\bibfnamefont {J.~M.}\ \bibnamefont
			{Renes}}, \ and\ \bibinfo {author} {\bibfnamefont {R.~W.}\ \bibnamefont
			{Spekkens}},\ }\href {\doibase 10.1103/PhysRevLett.111.250404} {\bibfield
		{journal} {\bibinfo  {journal} {Phys. Rev. Lett.}\ }\textbf {\bibinfo
			{volume} {111}},\ \bibinfo {pages} {250404} (\bibinfo {year}
		{2013})}\BibitemShut {NoStop}%
	\bibitem [{\citenamefont {Lostaglio}\ \emph {et~al.}(2015)\citenamefont
		{Lostaglio}, \citenamefont {Korzekwa}, \citenamefont {Jennings},\ and\
		\citenamefont {Rudolph}}]{Thermodynamics2}%
	\BibitemOpen
	\bibfield  {author} {\bibinfo {author} {\bibfnamefont {M.}~\bibnamefont
			{Lostaglio}}, \bibinfo {author} {\bibfnamefont {K.}~\bibnamefont {Korzekwa}},
		\bibinfo {author} {\bibfnamefont {D.}~\bibnamefont {Jennings}}, \ and\
		\bibinfo {author} {\bibfnamefont {T.}~\bibnamefont {Rudolph}},\ }\href
	{\doibase 10.1103/PhysRevX.5.021001} {\bibfield  {journal} {\bibinfo
			{journal} {Phys. Rev. X}\ }\textbf {\bibinfo {volume} {5}},\ \bibinfo {pages}
		{021001} (\bibinfo {year} {2015})}\BibitemShut {NoStop}%
	\bibitem [{\citenamefont {Streltsov}\ \emph
		{et~al.}(2017{\natexlab{a}})\citenamefont {Streltsov}, \citenamefont
		{Adesso},\ and\ \citenamefont {Plenio}}]{streltsov2016quantum}%
	\BibitemOpen
	\bibfield  {author} {\bibinfo {author} {\bibfnamefont {A.}~\bibnamefont
			{Streltsov}}, \bibinfo {author} {\bibfnamefont {G.}~\bibnamefont {Adesso}}, \
		and\ \bibinfo {author} {\bibfnamefont {M.~B.}\ \bibnamefont {Plenio}},\
	}\href@noop {} {\bibfield  {journal} {\bibinfo  {journal} {Rev. Mod. Phys.}\
		}\textbf {\bibinfo {volume} {89}},\ \bibinfo {pages} {041003} (\bibinfo
		{year} {2017}{\natexlab{a}})}\BibitemShut {NoStop}%
	\bibitem [{\citenamefont {Baumgratz}\ \emph {et~al.}(2014)\citenamefont
		{Baumgratz}, \citenamefont {Cramer},\ and\ \citenamefont
		{Plenio}}]{QuantifyingCoherence}%
	\BibitemOpen
	\bibfield  {author} {\bibinfo {author} {\bibfnamefont {T.}~\bibnamefont
			{Baumgratz}}, \bibinfo {author} {\bibfnamefont {M.}~\bibnamefont {Cramer}}, \
		and\ \bibinfo {author} {\bibfnamefont {M.~B.}\ \bibnamefont {Plenio}},\
	}\href {\doibase 10.1103/PhysRevLett.113.140401} {\bibfield  {journal}
		{\bibinfo  {journal} {Phys. Rev. Lett.}\ }\textbf {\bibinfo {volume} {113}},\
		\bibinfo {pages} {140401} (\bibinfo {year} {2014})}\BibitemShut {NoStop}%
	\bibitem [{\citenamefont {Girolami}(2014)}]{Giro14observable}%
	\BibitemOpen
	\bibfield  {author} {\bibinfo {author} {\bibfnamefont {D.}~\bibnamefont
			{Girolami}},\ }\href {\doibase 10.1103/PhysRevLett.113.170401} {\bibfield
		{journal} {\bibinfo  {journal} {Phys. Rev. Lett.}\ }\textbf {\bibinfo
			{volume} {113}},\ \bibinfo {pages} {170401} (\bibinfo {year}
		{2014})}\BibitemShut {NoStop}%
	\bibitem [{\citenamefont {Streltsov}\ \emph
		{et~al.}(2015{\natexlab{a}})\citenamefont {Streltsov}, \citenamefont {Singh},
		\citenamefont {Dhar}, \citenamefont {Bera},\ and\ \citenamefont
		{Adesso}}]{MeasuringQuantumCoherencewithEntanglement}%
	\BibitemOpen
	\bibfield  {author} {\bibinfo {author} {\bibfnamefont {A.}~\bibnamefont
			{Streltsov}}, \bibinfo {author} {\bibfnamefont {U.}~\bibnamefont {Singh}},
		\bibinfo {author} {\bibfnamefont {H.~S.}\ \bibnamefont {Dhar}}, \bibinfo
		{author} {\bibfnamefont {M.~N.}\ \bibnamefont {Bera}}, \ and\ \bibinfo
		{author} {\bibfnamefont {G.}~\bibnamefont {Adesso}},\ }\href {\doibase
		10.1103/PhysRevLett.115.020403} {\bibfield  {journal} {\bibinfo  {journal}
			{Phys. Rev. Lett.}\ }\textbf {\bibinfo {volume} {115}},\ \bibinfo {pages}
		{020403} (\bibinfo {year} {2015}{\natexlab{a}})}\BibitemShut {NoStop}%
	\bibitem [{\citenamefont {Winter}\ and\ \citenamefont
		{Yang}(2016)}]{OperationalResourceTheoryofCoherence}%
	\BibitemOpen
	\bibfield  {author} {\bibinfo {author} {\bibfnamefont {A.}~\bibnamefont
			{Winter}}\ and\ \bibinfo {author} {\bibfnamefont {D.}~\bibnamefont {Yang}},\
	}\href {\doibase 10.1103/PhysRevLett.116.120404} {\bibfield  {journal}
		{\bibinfo  {journal} {Phys. Rev. Lett.}\ }\textbf {\bibinfo {volume} {116}},\
		\bibinfo {pages} {120404} (\bibinfo {year} {2016})}\BibitemShut {NoStop}%
	\bibitem [{\citenamefont {Yuan}\ \emph {et~al.}(2015)\citenamefont {Yuan},
		\citenamefont {Zhou}, \citenamefont {Cao},\ and\ \citenamefont
		{Ma}}]{Yuan15intrinsic}%
	\BibitemOpen
	\bibfield  {author} {\bibinfo {author} {\bibfnamefont {X.}~\bibnamefont
			{Yuan}}, \bibinfo {author} {\bibfnamefont {H.}~\bibnamefont {Zhou}}, \bibinfo
		{author} {\bibfnamefont {Z.}~\bibnamefont {Cao}}, \ and\ \bibinfo {author}
		{\bibfnamefont {X.}~\bibnamefont {Ma}},\ }\href {\doibase
		10.1103/PhysRevA.92.022124} {\bibfield  {journal} {\bibinfo  {journal} {Phys.
				Rev. A}\ }\textbf {\bibinfo {volume} {92}},\ \bibinfo {pages} {022124}
		(\bibinfo {year} {2015})}\BibitemShut {NoStop}%
	\bibitem [{\citenamefont {Bromley}\ \emph {et~al.}(2015)\citenamefont
		{Bromley}, \citenamefont {Cianciaruso},\ and\ \citenamefont
		{Adesso}}]{FrozenQuantumCoherence}%
	\BibitemOpen
	\bibfield  {author} {\bibinfo {author} {\bibfnamefont {T.~R.}\ \bibnamefont
			{Bromley}}, \bibinfo {author} {\bibfnamefont {M.}~\bibnamefont
			{Cianciaruso}}, \ and\ \bibinfo {author} {\bibfnamefont {G.}~\bibnamefont
			{Adesso}},\ }\href {\doibase 10.1103/PhysRevLett.114.210401} {\bibfield
		{journal} {\bibinfo  {journal} {Phys. Rev. Lett.}\ }\textbf {\bibinfo
			{volume} {114}},\ \bibinfo {pages} {210401} (\bibinfo {year}
		{2015})}\BibitemShut {NoStop}%
	\bibitem [{\citenamefont {Silva}\ \emph {et~al.}(2016)\citenamefont {Silva},
		\citenamefont {Souza}, \citenamefont {Bromley}, \citenamefont {Cianciaruso},
		\citenamefont {Marx}, \citenamefont {Sarthour}, \citenamefont {Oliveira},
		\citenamefont {L.~Franco}, \citenamefont {Glaser}, \citenamefont {deAzevedo},
		\citenamefont {Soares-Pinto},\ and\ \citenamefont
		{Adesso}}]{TimeInvariantCoherenceNMR}%
	\BibitemOpen
	\bibfield  {author} {\bibinfo {author} {\bibfnamefont {I.~A.}\ \bibnamefont
			{Silva}}, \bibinfo {author} {\bibfnamefont {A.~M.}\ \bibnamefont {Souza}},
		\bibinfo {author} {\bibfnamefont {T.~R.}\ \bibnamefont {Bromley}}, \bibinfo
		{author} {\bibfnamefont {M.}~\bibnamefont {Cianciaruso}}, \bibinfo {author}
		{\bibfnamefont {R.}~\bibnamefont {Marx}}, \bibinfo {author} {\bibfnamefont
			{R.~S.}\ \bibnamefont {Sarthour}}, \bibinfo {author} {\bibfnamefont {I.~S.}\
			\bibnamefont {Oliveira}}, \bibinfo {author} {\bibfnamefont {R.}~\bibnamefont
			{L.~Franco}}, \bibinfo {author} {\bibfnamefont {S.~J.}\ \bibnamefont
			{Glaser}}, \bibinfo {author} {\bibfnamefont {E.~R.}\ \bibnamefont
			{deAzevedo}}, \bibinfo {author} {\bibfnamefont {D.~O.}\ \bibnamefont
			{Soares-Pinto}}, \ and\ \bibinfo {author} {\bibfnamefont {G.}~\bibnamefont
			{Adesso}},\ }\href {\doibase 10.1103/PhysRevLett.117.160402} {\bibfield
		{journal} {\bibinfo  {journal} {Phys. Rev. Lett.}\ }\textbf {\bibinfo
			{volume} {117}},\ \bibinfo {pages} {160402} (\bibinfo {year}
		{2016})}\BibitemShut {NoStop}%
	\bibitem [{\citenamefont {Huang}\ and\ \citenamefont
		{Situ}(2017)}]{huang2017non}%
	\BibitemOpen
	\bibfield  {author} {\bibinfo {author} {\bibfnamefont {Z.}~\bibnamefont
			{Huang}}\ and\ \bibinfo {author} {\bibfnamefont {H.}~\bibnamefont {Situ}},\
	}\href@noop {} {\bibfield  {journal} {\bibinfo  {journal} {Quantum Inf.
				Process.}\ }\textbf {\bibinfo {volume} {16}},\ \bibinfo {pages} {222}
		(\bibinfo {year} {2017})}\BibitemShut {NoStop}%
	\bibitem [{\citenamefont {Qin}\ \emph {et~al.}(2018)\citenamefont {Qin},
		\citenamefont {Ren},\ and\ \citenamefont {Zhang}}]{qin2018dynamics}%
	\BibitemOpen
	\bibfield  {author} {\bibinfo {author} {\bibfnamefont {M.}~\bibnamefont
			{Qin}}, \bibinfo {author} {\bibfnamefont {Z.}~\bibnamefont {Ren}}, \ and\
		\bibinfo {author} {\bibfnamefont {X.}~\bibnamefont {Zhang}},\ }\href@noop {}
	{\bibfield  {journal} {\bibinfo  {journal} {Phys. Rev. A}\ }\textbf {\bibinfo
			{volume} {98}},\ \bibinfo {pages} {012303} (\bibinfo {year}
		{2018})}\BibitemShut {NoStop}%
	\bibitem [{\citenamefont {Lostaglio}\ \emph {et~al.}(2017)\citenamefont
		{Lostaglio}, \citenamefont {Korzekwa},\ and\ \citenamefont
		{Milne}}]{lostaglio2017markovian}%
	\BibitemOpen
	\bibfield  {author} {\bibinfo {author} {\bibfnamefont {M.}~\bibnamefont
			{Lostaglio}}, \bibinfo {author} {\bibfnamefont {K.}~\bibnamefont {Korzekwa}},
		\ and\ \bibinfo {author} {\bibfnamefont {A.}~\bibnamefont {Milne}},\
	}\href@noop {} {\bibfield  {journal} {\bibinfo  {journal} {Phys. Rev. A}\
		}\textbf {\bibinfo {volume} {96}},\ \bibinfo {pages} {032109} (\bibinfo
		{year} {2017})}\BibitemShut {NoStop}%
	\bibitem [{\citenamefont {Man}\ \emph {et~al.}(2015)\citenamefont {Man},
		\citenamefont {Xia},\ and\ \citenamefont {Franco}}]{man2015cavity}%
	\BibitemOpen
	\bibfield  {author} {\bibinfo {author} {\bibfnamefont {Z.-X.}\ \bibnamefont
			{Man}}, \bibinfo {author} {\bibfnamefont {Y.-J.}\ \bibnamefont {Xia}}, \ and\
		\bibinfo {author} {\bibfnamefont {R.~L.}\ \bibnamefont {Franco}},\
	}\href@noop {} {\bibfield  {journal} {\bibinfo  {journal} {Sci. Rep.}\
		}\textbf {\bibinfo {volume} {5}},\ \bibinfo {pages} {13843} (\bibinfo {year}
		{2015})}\BibitemShut {NoStop}%
	\bibitem [{\citenamefont {Breuer}\ and\ \citenamefont
		{Petruccione}(2002)}]{breuer2002theory}%
	\BibitemOpen
	\bibfield  {author} {\bibinfo {author} {\bibfnamefont {H.-P.}\ \bibnamefont
			{Breuer}}\ and\ \bibinfo {author} {\bibfnamefont {F.}~\bibnamefont
			{Petruccione}},\ }\href@noop {} {\emph {\bibinfo {title} {The {T}heory of
				{O}pen {Q}uantum {S}ystems}}}\ (\bibinfo  {publisher} {Oxford University
		Press on Demand},\ \bibinfo {year} {2002})\BibitemShut {NoStop}%
	\bibitem [{\citenamefont {Weiss}(2012)}]{weiss2012quantum}%
	\BibitemOpen
	\bibfield  {author} {\bibinfo {author} {\bibfnamefont {U.}~\bibnamefont
			{Weiss}},\ }\href@noop {} {\emph {\bibinfo {title} {Quantum {D}issipative
				{S}ystems}}},\ Vol.~\bibinfo {volume} {13}\ (\bibinfo  {publisher} {World
		scientific},\ \bibinfo {year} {2012})\BibitemShut {NoStop}%
	\bibitem [{\citenamefont {Zhang}\ \emph {et~al.}(2012)\citenamefont {Zhang},
		\citenamefont {Lo}, \citenamefont {Xiong}, \citenamefont {Tu},\ and\
		\citenamefont {Nori}}]{PhysRevLettGeneraldynamics}%
	\BibitemOpen
	\bibfield  {author} {\bibinfo {author} {\bibfnamefont {W.-M.}\ \bibnamefont
			{Zhang}}, \bibinfo {author} {\bibfnamefont {P.-Y.}\ \bibnamefont {Lo}},
		\bibinfo {author} {\bibfnamefont {H.-N.}\ \bibnamefont {Xiong}}, \bibinfo
		{author} {\bibfnamefont {M.~W.-Y.}\ \bibnamefont {Tu}}, \ and\ \bibinfo
		{author} {\bibfnamefont {F.}~\bibnamefont {Nori}},\ }\href {\doibase
		10.1103/PhysRevLett.109.170402} {\bibfield  {journal} {\bibinfo  {journal}
			{Phys. Rev. Lett.}\ }\textbf {\bibinfo {volume} {109}},\ \bibinfo {pages}
		{170402} (\bibinfo {year} {2012})}\BibitemShut {NoStop}%
	\bibitem [{\citenamefont {Chen}\ \emph {et~al.}(2009)\citenamefont {Chen},
		\citenamefont {Chen}, \citenamefont {Liao}, \citenamefont {Lambert},\ and\
		\citenamefont {Nori}}]{PhysRevBquantumdots}%
	\BibitemOpen
	\bibfield  {author} {\bibinfo {author} {\bibfnamefont {Y.-N.}\ \bibnamefont
			{Chen}}, \bibinfo {author} {\bibfnamefont {G.-Y.}\ \bibnamefont {Chen}},
		\bibinfo {author} {\bibfnamefont {Y.-Y.}\ \bibnamefont {Liao}}, \bibinfo
		{author} {\bibfnamefont {N.}~\bibnamefont {Lambert}}, \ and\ \bibinfo
		{author} {\bibfnamefont {F.}~\bibnamefont {Nori}},\ }\href {\doibase
		10.1103/PhysRevB.79.245312} {\bibfield  {journal} {\bibinfo  {journal} {Phys.
				Rev. B}\ }\textbf {\bibinfo {volume} {79}},\ \bibinfo {pages} {245312}
		(\bibinfo {year} {2009})}\BibitemShut {NoStop}%
	\bibitem [{\citenamefont {Yin}\ \emph {et~al.}(2012)\citenamefont {Yin},
		\citenamefont {Ma}, \citenamefont {Wang},\ and\ \citenamefont
		{Nori}}]{PhysRevASpinNonMar}%
	\BibitemOpen
	\bibfield  {author} {\bibinfo {author} {\bibfnamefont {X.}~\bibnamefont
			{Yin}}, \bibinfo {author} {\bibfnamefont {J.}~\bibnamefont {Ma}}, \bibinfo
		{author} {\bibfnamefont {X.}~\bibnamefont {Wang}}, \ and\ \bibinfo {author}
		{\bibfnamefont {F.}~\bibnamefont {Nori}},\ }\href {\doibase
		10.1103/PhysRevA.86.012308} {\bibfield  {journal} {\bibinfo  {journal} {Phys.
				Rev. A}\ }\textbf {\bibinfo {volume} {86}},\ \bibinfo {pages} {012308}
		(\bibinfo {year} {2012})}\BibitemShut {NoStop}%
	\bibitem [{\citenamefont {Zhang}\ \emph {et~al.}(2013)\citenamefont {Zhang},
		\citenamefont {Liu}, \citenamefont {Wu}, \citenamefont {Jacobs},\ and\
		\citenamefont {Nori}}]{PhysRevAnonMarinputoutput}%
	\BibitemOpen
	\bibfield  {author} {\bibinfo {author} {\bibfnamefont {J.}~\bibnamefont
			{Zhang}}, \bibinfo {author} {\bibfnamefont {Y.-X.}\ \bibnamefont {Liu}},
		\bibinfo {author} {\bibfnamefont {R.-B.}\ \bibnamefont {Wu}}, \bibinfo
		{author} {\bibfnamefont {K.}~\bibnamefont {Jacobs}}, \ and\ \bibinfo {author}
		{\bibfnamefont {F.}~\bibnamefont {Nori}},\ }\href {\doibase
		10.1103/PhysRevA.87.032117} {\bibfield  {journal} {\bibinfo  {journal} {Phys.
				Rev. A}\ }\textbf {\bibinfo {volume} {87}},\ \bibinfo {pages} {032117}
		(\bibinfo {year} {2013})}\BibitemShut {NoStop}%
	\bibitem [{\citenamefont {Chen}\ \emph {et~al.}(2015)\citenamefont {Chen},
		\citenamefont {Lambert}, \citenamefont {Cheng}, \citenamefont {Chen},\ and\
		\citenamefont {Nori}}]{chen2015using}%
	\BibitemOpen
	\bibfield  {author} {\bibinfo {author} {\bibfnamefont {H.-B.}\ \bibnamefont
			{Chen}}, \bibinfo {author} {\bibfnamefont {N.}~\bibnamefont {Lambert}},
		\bibinfo {author} {\bibfnamefont {Y.-C.}\ \bibnamefont {Cheng}}, \bibinfo
		{author} {\bibfnamefont {Y.-N.}\ \bibnamefont {Chen}}, \ and\ \bibinfo
		{author} {\bibfnamefont {F.}~\bibnamefont {Nori}},\ }\href@noop {} {\bibfield
		{journal} {\bibinfo  {journal} {Sci. Rep.}\ }\textbf {\bibinfo {volume}
			{5}},\ \bibinfo {pages} {12753} (\bibinfo {year} {2015})}\BibitemShut
	{NoStop}%
	\bibitem [{\citenamefont {Xiong}\ \emph {et~al.}(2015)\citenamefont {Xiong},
		\citenamefont {Lo}, \citenamefont {Zhang}, \citenamefont {Nori} \emph
		{et~al.}}]{xiong2015non}%
	\BibitemOpen
	\bibfield  {author} {\bibinfo {author} {\bibfnamefont {H.-N.}\ \bibnamefont
			{Xiong}}, \bibinfo {author} {\bibfnamefont {P.-Y.}\ \bibnamefont {Lo}},
		\bibinfo {author} {\bibfnamefont {W.-M.}\ \bibnamefont {Zhang}}, \bibinfo
		{author} {\bibfnamefont {F.}~\bibnamefont {Nori}},  \emph {et~al.},\
	}\href@noop {} {\bibfield  {journal} {\bibinfo  {journal} {Sci. Rep.}\
		}\textbf {\bibinfo {volume} {5}},\ \bibinfo {pages} {13353} (\bibinfo {year}
		{2015})}\BibitemShut {NoStop}%
	\bibitem [{\citenamefont {Pollock}\ \emph {et~al.}(2018)\citenamefont
		{Pollock}, \citenamefont {Rodr{\'\i}guez-Rosario}, \citenamefont
		{Frauenheim}, \citenamefont {Paternostro},\ and\ \citenamefont
		{Modi}}]{pollock2018operational}%
	\BibitemOpen
	\bibfield  {author} {\bibinfo {author} {\bibfnamefont {F.~A.}\ \bibnamefont
			{Pollock}}, \bibinfo {author} {\bibfnamefont {C.}~\bibnamefont
			{Rodr{\'\i}guez-Rosario}}, \bibinfo {author} {\bibfnamefont {T.}~\bibnamefont
			{Frauenheim}}, \bibinfo {author} {\bibfnamefont {M.}~\bibnamefont
			{Paternostro}}, \ and\ \bibinfo {author} {\bibfnamefont {K.}~\bibnamefont
			{Modi}},\ }\href@noop {} {\bibfield  {journal} {\bibinfo  {journal} {Phys.
				Rev. Lett.}\ }\textbf {\bibinfo {volume} {120}},\ \bibinfo {pages} {040405}
		(\bibinfo {year} {2018})}\BibitemShut {NoStop}%
	\bibitem [{\citenamefont {Gorini}\ \emph {et~al.}(1976)\citenamefont {Gorini},
		\citenamefont {Kossakowski},\ and\ \citenamefont
		{Sudarshan}}]{gorini1976completely}%
	\BibitemOpen
	\bibfield  {author} {\bibinfo {author} {\bibfnamefont {V.}~\bibnamefont
			{Gorini}}, \bibinfo {author} {\bibfnamefont {A.}~\bibnamefont {Kossakowski}},
		\ and\ \bibinfo {author} {\bibfnamefont {E.~C.~G.}\ \bibnamefont
			{Sudarshan}},\ }\href@noop {} {\bibfield  {journal} {\bibinfo  {journal} {J.
				Math. Phys.}\ }\textbf {\bibinfo {volume} {17}},\ \bibinfo {pages} {821}
		(\bibinfo {year} {1976})}\BibitemShut {NoStop}%
	\bibitem [{\citenamefont {Lindblad}(1976)}]{lindblad1976generators}%
	\BibitemOpen
	\bibfield  {author} {\bibinfo {author} {\bibfnamefont {G.}~\bibnamefont
			{Lindblad}},\ }\href@noop {} {\bibfield  {journal} {\bibinfo  {journal}
			{Commun. Math. Phys.}\ }\textbf {\bibinfo {volume} {48}},\ \bibinfo {pages}
		{119} (\bibinfo {year} {1976})}\BibitemShut {NoStop}%
	\bibitem [{\citenamefont {Breuer}\ \emph {et~al.}(2009)\citenamefont {Breuer},
		\citenamefont {Laine},\ and\ \citenamefont {Piilo}}]{MeasureBreuer2009}%
	\BibitemOpen
	\bibfield  {author} {\bibinfo {author} {\bibfnamefont {H.-P.}\ \bibnamefont
			{Breuer}}, \bibinfo {author} {\bibfnamefont {E.-M.}\ \bibnamefont {Laine}}, \
		and\ \bibinfo {author} {\bibfnamefont {J.}~\bibnamefont {Piilo}},\ }\href
	{\doibase 10.1103/PhysRevLett.103.210401} {\bibfield  {journal} {\bibinfo
			{journal} {Phys. Rev. Lett.}\ }\textbf {\bibinfo {volume} {103}},\ \bibinfo
		{pages} {210401} (\bibinfo {year} {2009})}\BibitemShut {NoStop}%
	\bibitem [{\citenamefont {Rivas}\ \emph {et~al.}(2010)\citenamefont {Rivas},
		\citenamefont {Huelga},\ and\ \citenamefont
		{Plenio}}]{EntanglementRivas2010}%
	\BibitemOpen
	\bibfield  {author} {\bibinfo {author} {\bibfnamefont {A.}~\bibnamefont
			{Rivas}}, \bibinfo {author} {\bibfnamefont {S.~F.}\ \bibnamefont {Huelga}}, \
		and\ \bibinfo {author} {\bibfnamefont {M.~B.}\ \bibnamefont {Plenio}},\
	}\href {\doibase 10.1103/PhysRevLett.105.050403} {\bibfield  {journal}
		{\bibinfo  {journal} {Phys. Rev. Lett.}\ }\textbf {\bibinfo {volume} {105}},\
		\bibinfo {pages} {050403} (\bibinfo {year} {2010})}\BibitemShut {NoStop}%
	\bibitem [{\citenamefont {Lu}\ \emph {et~al.}(2010)\citenamefont {Lu},
		\citenamefont {Wang},\ and\ \citenamefont {Sun}}]{SunCPQFI2010}%
	\BibitemOpen
	\bibfield  {author} {\bibinfo {author} {\bibfnamefont {X.-M.}\ \bibnamefont
			{Lu}}, \bibinfo {author} {\bibfnamefont {X.}~\bibnamefont {Wang}}, \ and\
		\bibinfo {author} {\bibfnamefont {C.~P.}\ \bibnamefont {Sun}},\ }\href
	{\doibase 10.1103/PhysRevA.82.042103} {\bibfield  {journal} {\bibinfo
			{journal} {Phys. Rev. A}\ }\textbf {\bibinfo {volume} {82}},\ \bibinfo
		{pages} {042103} (\bibinfo {year} {2010})}\BibitemShut {NoStop}%
	\bibitem [{\citenamefont {Song}\ \emph {et~al.}(2015)\citenamefont {Song},
		\citenamefont {Luo},\ and\ \citenamefont {Hong}}]{LuoShunlongQFI2010}%
	\BibitemOpen
	\bibfield  {author} {\bibinfo {author} {\bibfnamefont {H.}~\bibnamefont
			{Song}}, \bibinfo {author} {\bibfnamefont {S.}~\bibnamefont {Luo}}, \ and\
		\bibinfo {author} {\bibfnamefont {Y.}~\bibnamefont {Hong}},\ }\href {\doibase
		10.1103/PhysRevA.91.042110} {\bibfield  {journal} {\bibinfo  {journal} {Phys.
				Rev. A}\ }\textbf {\bibinfo {volume} {91}},\ \bibinfo {pages} {042110}
		(\bibinfo {year} {2015})}\BibitemShut {NoStop}%
	\bibitem [{\citenamefont {Rajagopal}\ \emph {et~al.}(2010)\citenamefont
		{Rajagopal}, \citenamefont {Usha~Devi},\ and\ \citenamefont
		{Rendell}}]{Rajagopalfidelity2010}%
	\BibitemOpen
	\bibfield  {author} {\bibinfo {author} {\bibfnamefont {A.~K.}\ \bibnamefont
			{Rajagopal}}, \bibinfo {author} {\bibfnamefont {A.~R.}\ \bibnamefont
			{Usha~Devi}}, \ and\ \bibinfo {author} {\bibfnamefont {R.~W.}\ \bibnamefont
			{Rendell}},\ }\href {\doibase 10.1103/PhysRevA.82.042107} {\bibfield
		{journal} {\bibinfo  {journal} {Phys. Rev. A}\ }\textbf {\bibinfo {volume}
			{82}},\ \bibinfo {pages} {042107} (\bibinfo {year} {2010})}\BibitemShut
	{NoStop}%
	\bibitem [{\citenamefont {Luo}\ \emph {et~al.}(2012)\citenamefont {Luo},
		\citenamefont {Fu},\ and\ \citenamefont {Song}}]{correlations2012Luo}%
	\BibitemOpen
	\bibfield  {author} {\bibinfo {author} {\bibfnamefont {S.}~\bibnamefont
			{Luo}}, \bibinfo {author} {\bibfnamefont {S.}~\bibnamefont {Fu}}, \ and\
		\bibinfo {author} {\bibfnamefont {H.}~\bibnamefont {Song}},\ }\href {\doibase
		10.1103/PhysRevA.86.044101} {\bibfield  {journal} {\bibinfo  {journal} {Phys.
				Rev. A}\ }\textbf {\bibinfo {volume} {86}},\ \bibinfo {pages} {044101}
		(\bibinfo {year} {2012})}\BibitemShut {NoStop}%
	\bibitem [{\citenamefont {Bylicka}\ \emph {et~al.}(2014)\citenamefont
		{Bylicka}, \citenamefont {Chru{\'s}ci{\'n}ski},\ and\ \citenamefont
		{Maniscalco}}]{bylicka2014non}%
	\BibitemOpen
	\bibfield  {author} {\bibinfo {author} {\bibfnamefont {B.}~\bibnamefont
			{Bylicka}}, \bibinfo {author} {\bibfnamefont {D.}~\bibnamefont
			{Chru{\'s}ci{\'n}ski}}, \ and\ \bibinfo {author} {\bibfnamefont
			{S.}~\bibnamefont {Maniscalco}},\ }\href@noop {} {\bibfield  {journal}
		{\bibinfo  {journal} {Sci. Rep.}\ }\textbf {\bibinfo {volume} {4}},\ \bibinfo
		{pages} {5720} (\bibinfo {year} {2014})}\BibitemShut {NoStop}%
	\bibitem [{\citenamefont {Lorenzo}\ \emph {et~al.}(2013)\citenamefont
		{Lorenzo}, \citenamefont {Plastina},\ and\ \citenamefont
		{Paternostro}}]{Geometric2013Lorenzo}%
	\BibitemOpen
	\bibfield  {author} {\bibinfo {author} {\bibfnamefont {S.}~\bibnamefont
			{Lorenzo}}, \bibinfo {author} {\bibfnamefont {F.}~\bibnamefont {Plastina}}, \
		and\ \bibinfo {author} {\bibfnamefont {M.}~\bibnamefont {Paternostro}},\
	}\href {\doibase 10.1103/PhysRevA.88.020102} {\bibfield  {journal} {\bibinfo
			{journal} {Phys. Rev. A}\ }\textbf {\bibinfo {volume} {88}},\ \bibinfo
		{pages} {020102} (\bibinfo {year} {2013})}\BibitemShut {NoStop}%
	\bibitem [{\citenamefont {Bae}\ and\ \citenamefont {Chru\ifmmode \acute{s}\else
			\'{s}\fi{}ci\ifmmode~\acute{n}\else
			\'{n}\fi{}ski}(2016)}]{OperationalDivisibility2016Joonwoo}%
	\BibitemOpen
	\bibfield  {author} {\bibinfo {author} {\bibfnamefont {J.}~\bibnamefont
			{Bae}}\ and\ \bibinfo {author} {\bibfnamefont {D.}~\bibnamefont {Chru\ifmmode
				\acute{s}\else \'{s}\fi{}ci\ifmmode~\acute{n}\else \'{n}\fi{}ski}},\ }\href
	{\doibase 10.1103/PhysRevLett.117.050403} {\bibfield  {journal} {\bibinfo
			{journal} {Phys. Rev. Lett.}\ }\textbf {\bibinfo {volume} {117}},\ \bibinfo
		{pages} {050403} (\bibinfo {year} {2016})}\BibitemShut {NoStop}%
	\bibitem [{\citenamefont {Chen}\ \emph {et~al.}(2016)\citenamefont {Chen},
		\citenamefont {Lambert}, \citenamefont {Li}, \citenamefont {Miranowicz},
		\citenamefont {Chen},\ and\ \citenamefont
		{Nori}}]{PRL2016nonMarTemporalsteering}%
	\BibitemOpen
	\bibfield  {author} {\bibinfo {author} {\bibfnamefont {S.-L.}\ \bibnamefont
			{Chen}}, \bibinfo {author} {\bibfnamefont {N.}~\bibnamefont {Lambert}},
		\bibinfo {author} {\bibfnamefont {C.-M.}\ \bibnamefont {Li}}, \bibinfo
		{author} {\bibfnamefont {A.}~\bibnamefont {Miranowicz}}, \bibinfo {author}
		{\bibfnamefont {Y.-N.}\ \bibnamefont {Chen}}, \ and\ \bibinfo {author}
		{\bibfnamefont {F.}~\bibnamefont {Nori}},\ }\href {\doibase
		10.1103/PhysRevLett.116.020503} {\bibfield  {journal} {\bibinfo  {journal}
			{Phys. Rev. Lett.}\ }\textbf {\bibinfo {volume} {116}},\ \bibinfo {pages}
		{020503} (\bibinfo {year} {2016})}\BibitemShut {NoStop}%
	\bibitem [{\citenamefont {Strathearn}\ \emph {et~al.}(2018)\citenamefont
		{Strathearn}, \citenamefont {Kirton}, \citenamefont {Kilda}, \citenamefont
		{Keeling},\ and\ \citenamefont {Lovett}}]{strathearn2018efficient}%
	\BibitemOpen
	\bibfield  {author} {\bibinfo {author} {\bibfnamefont {A.}~\bibnamefont
			{Strathearn}}, \bibinfo {author} {\bibfnamefont {P.}~\bibnamefont {Kirton}},
		\bibinfo {author} {\bibfnamefont {D.}~\bibnamefont {Kilda}}, \bibinfo
		{author} {\bibfnamefont {J.}~\bibnamefont {Keeling}}, \ and\ \bibinfo
		{author} {\bibfnamefont {B.~W.}\ \bibnamefont {Lovett}},\ }\href@noop {}
	{\bibfield  {journal} {\bibinfo  {journal} {Nat. Commun.}\ }\textbf {\bibinfo
			{volume} {9}},\ \bibinfo {pages} {3322} (\bibinfo {year} {2018})}\BibitemShut
	{NoStop}%
	\bibitem [{\citenamefont {Ku}\ \emph {et~al.}(2016)\citenamefont {Ku},
		\citenamefont {Chen}, \citenamefont {Chen}, \citenamefont {Lambert},
		\citenamefont {Chen},\ and\ \citenamefont {Nori}}]{PhysRevA.94.062126}%
	\BibitemOpen
	\bibfield  {author} {\bibinfo {author} {\bibfnamefont {H.-Y.}\ \bibnamefont
			{Ku}}, \bibinfo {author} {\bibfnamefont {S.-L.}\ \bibnamefont {Chen}},
		\bibinfo {author} {\bibfnamefont {H.-B.}\ \bibnamefont {Chen}}, \bibinfo
		{author} {\bibfnamefont {N.}~\bibnamefont {Lambert}}, \bibinfo {author}
		{\bibfnamefont {Y.-N.}\ \bibnamefont {Chen}}, \ and\ \bibinfo {author}
		{\bibfnamefont {F.}~\bibnamefont {Nori}},\ }\href {\doibase
		10.1103/PhysRevA.94.062126} {\bibfield  {journal} {\bibinfo  {journal} {Phys.
				Rev. A}\ }\textbf {\bibinfo {volume} {94}},\ \bibinfo {pages} {062126}
		(\bibinfo {year} {2016})}\BibitemShut {NoStop}%
	\bibitem [{\citenamefont {Xiong}\ \emph {et~al.}(2017)\citenamefont {Xiong},
		\citenamefont {Zhang}, \citenamefont {Sun}, \citenamefont {Yu}, \citenamefont
		{Su}, \citenamefont {Xu}, \citenamefont {Jin}, \citenamefont {Xu},
		\citenamefont {Liu}, \citenamefont {Chen} \emph
		{et~al.}}]{xiong2017experimental}%
	\BibitemOpen
	\bibfield  {author} {\bibinfo {author} {\bibfnamefont {S.-J.}\ \bibnamefont
			{Xiong}}, \bibinfo {author} {\bibfnamefont {Y.}~\bibnamefont {Zhang}},
		\bibinfo {author} {\bibfnamefont {Z.}~\bibnamefont {Sun}}, \bibinfo {author}
		{\bibfnamefont {L.}~\bibnamefont {Yu}}, \bibinfo {author} {\bibfnamefont
			{Q.}~\bibnamefont {Su}}, \bibinfo {author} {\bibfnamefont {X.-Q.}\
			\bibnamefont {Xu}}, \bibinfo {author} {\bibfnamefont {J.-S.}\ \bibnamefont
			{Jin}}, \bibinfo {author} {\bibfnamefont {Q.}~\bibnamefont {Xu}}, \bibinfo
		{author} {\bibfnamefont {J.-M.}\ \bibnamefont {Liu}}, \bibinfo {author}
		{\bibfnamefont {K.}~\bibnamefont {Chen}},  \emph {et~al.},\ }\href@noop {}
	{\bibfield  {journal} {\bibinfo  {journal} {Optica}\ }\textbf {\bibinfo
			{volume} {4}},\ \bibinfo {pages} {1065} (\bibinfo {year} {2017})}\BibitemShut
	{NoStop}%
	\bibitem [{\citenamefont {Li}\ \emph {et~al.}(2015)\citenamefont {Li},
		\citenamefont {Chen}, \citenamefont {Lambert}, \citenamefont {Chiu},\ and\
		\citenamefont {Nori}}]{PhysRevA.92.062310}%
	\BibitemOpen
	\bibfield  {author} {\bibinfo {author} {\bibfnamefont {C.-M.}\ \bibnamefont
			{Li}}, \bibinfo {author} {\bibfnamefont {Y.-N.}\ \bibnamefont {Chen}},
		\bibinfo {author} {\bibfnamefont {N.}~\bibnamefont {Lambert}}, \bibinfo
		{author} {\bibfnamefont {C.-Y.}\ \bibnamefont {Chiu}}, \ and\ \bibinfo
		{author} {\bibfnamefont {F.}~\bibnamefont {Nori}},\ }\href {\doibase
		10.1103/PhysRevA.92.062310} {\bibfield  {journal} {\bibinfo  {journal} {Phys.
				Rev. A}\ }\textbf {\bibinfo {volume} {92}},\ \bibinfo {pages} {062310}
		(\bibinfo {year} {2015})}\BibitemShut {NoStop}%
	\bibitem [{\citenamefont {He}\ \emph {et~al.}(2017)\citenamefont {He},
		\citenamefont {Zeng}, \citenamefont {Li}, \citenamefont {Wang},\ and\
		\citenamefont {Yao}}]{he2017non}%
	\BibitemOpen
	\bibfield  {author} {\bibinfo {author} {\bibfnamefont {Z.}~\bibnamefont
			{He}}, \bibinfo {author} {\bibfnamefont {H.-S.}\ \bibnamefont {Zeng}},
		\bibinfo {author} {\bibfnamefont {Y.}~\bibnamefont {Li}}, \bibinfo {author}
		{\bibfnamefont {Q.}~\bibnamefont {Wang}}, \ and\ \bibinfo {author}
		{\bibfnamefont {C.}~\bibnamefont {Yao}},\ }\href@noop {} {\bibfield
		{journal} {\bibinfo  {journal} {Phys. Rev. A}\ }\textbf {\bibinfo {volume}
			{96}},\ \bibinfo {pages} {022106} (\bibinfo {year} {2017})}\BibitemShut
	{NoStop}%
	\bibitem [{\citenamefont {Radhakrishnan}\ \emph {et~al.}(2017)\citenamefont
		{Radhakrishnan}, \citenamefont {Chen}, \citenamefont {Jambulingam},
		\citenamefont {Byrnes}, \citenamefont {Ali} \emph
		{et~al.}}]{radhakrishnan2017time}%
	\BibitemOpen
	\bibfield  {author} {\bibinfo {author} {\bibfnamefont {C.}~\bibnamefont
			{Radhakrishnan}}, \bibinfo {author} {\bibfnamefont {P.-W.}\ \bibnamefont
			{Chen}}, \bibinfo {author} {\bibfnamefont {S.}~\bibnamefont {Jambulingam}},
		\bibinfo {author} {\bibfnamefont {T.}~\bibnamefont {Byrnes}}, \bibinfo
		{author} {\bibfnamefont {M.}~\bibnamefont {Ali}},  \emph {et~al.},\
	}\href@noop {} {\bibfield  {journal} {\bibinfo  {journal} {arXiv:1711.03299}\
		} (\bibinfo {year} {2017})}\BibitemShut {NoStop}%
	\bibitem [{\citenamefont {Chanda}\ and\ \citenamefont
		{Bhattacharya}(2016)}]{chanda2016delineating}%
	\BibitemOpen
	\bibfield  {author} {\bibinfo {author} {\bibfnamefont {T.}~\bibnamefont
			{Chanda}}\ and\ \bibinfo {author} {\bibfnamefont {S.}~\bibnamefont
			{Bhattacharya}},\ }\href@noop {} {\bibfield  {journal} {\bibinfo  {journal}
			{Ann. Phys.}\ }\textbf {\bibinfo {volume} {366}},\ \bibinfo {pages} {1}
		(\bibinfo {year} {2016})}\BibitemShut {NoStop}%
	\bibitem [{\citenamefont {Mirafzali}\ and\ \citenamefont
		{Baghshahi}(2019)}]{mirafzali2019non}%
	\BibitemOpen
	\bibfield  {author} {\bibinfo {author} {\bibfnamefont {S.~Y.}\ \bibnamefont
			{Mirafzali}}\ and\ \bibinfo {author} {\bibfnamefont {H.~R.}\ \bibnamefont
			{Baghshahi}},\ }\href@noop {} {\bibfield  {journal} {\bibinfo  {journal}
			{Physica A: Statistical Mechanics and its Applications}\ }\textbf {\bibinfo
			{volume} {514}},\ \bibinfo {pages} {274} (\bibinfo {year}
		{2019})}\BibitemShut {NoStop}%
	\bibitem [{\citenamefont {Zhang}\ \emph {et~al.}(2015)\citenamefont {Zhang},
		\citenamefont {Han}, \citenamefont {Xia}, \citenamefont {Yu},\ and\
		\citenamefont {Fan}}]{zhang2015role}%
	\BibitemOpen
	\bibfield  {author} {\bibinfo {author} {\bibfnamefont {Y.-J.}\ \bibnamefont
			{Zhang}}, \bibinfo {author} {\bibfnamefont {W.}~\bibnamefont {Han}}, \bibinfo
		{author} {\bibfnamefont {Y.-J.}\ \bibnamefont {Xia}}, \bibinfo {author}
		{\bibfnamefont {Y.-M.}\ \bibnamefont {Yu}}, \ and\ \bibinfo {author}
		{\bibfnamefont {H.}~\bibnamefont {Fan}},\ }\href@noop {} {\bibfield
		{journal} {\bibinfo  {journal} {Sci. Rep.}\ }\textbf {\bibinfo {volume}
			{5}},\ \bibinfo {pages} {13359} (\bibinfo {year} {2015})}\BibitemShut
	{NoStop}%
	\bibitem [{\citenamefont {Addis}\ \emph {et~al.}(2014)\citenamefont {Addis},
		\citenamefont {Brebner}, \citenamefont {Haikka},\ and\ \citenamefont
		{Maniscalco}}]{coherencetrappingPRA}%
	\BibitemOpen
	\bibfield  {author} {\bibinfo {author} {\bibfnamefont {C.}~\bibnamefont
			{Addis}}, \bibinfo {author} {\bibfnamefont {G.}~\bibnamefont {Brebner}},
		\bibinfo {author} {\bibfnamefont {P.}~\bibnamefont {Haikka}}, \ and\ \bibinfo
		{author} {\bibfnamefont {S.}~\bibnamefont {Maniscalco}},\ }\href {\doibase
		10.1103/PhysRevA.89.024101} {\bibfield  {journal} {\bibinfo  {journal} {Phys.
				Rev. A}\ }\textbf {\bibinfo {volume} {89}},\ \bibinfo {pages} {024101}
		(\bibinfo {year} {2014})}\BibitemShut {NoStop}%
	\bibitem [{\citenamefont {Passos}\ \emph {et~al.}(2018)\citenamefont {Passos},
		\citenamefont {Obando}, \citenamefont {Balthazar}, \citenamefont {Paula},
		\citenamefont {Huguenin},\ and\ \citenamefont {Sarandy}}]{passos2018non}%
	\BibitemOpen
	\bibfield  {author} {\bibinfo {author} {\bibfnamefont {M.~H.~M.}\
			\bibnamefont {Passos}}, \bibinfo {author} {\bibfnamefont {P.~C.}\
			\bibnamefont {Obando}}, \bibinfo {author} {\bibfnamefont {W.~F.}\
			\bibnamefont {Balthazar}}, \bibinfo {author} {\bibfnamefont {F.~M.}\
			\bibnamefont {Paula}}, \bibinfo {author} {\bibfnamefont {J.~A.~O.}\
			\bibnamefont {Huguenin}}, \ and\ \bibinfo {author} {\bibfnamefont {M.~S.}\
			\bibnamefont {Sarandy}},\ }\href@noop {} {\bibfield  {journal} {\bibinfo
			{journal} {arXiv:1807.05378}\ } (\bibinfo {year} {2018})}\BibitemShut
	{NoStop}%
	\bibitem [{\citenamefont {Man}\ \emph {et~al.}(2018)\citenamefont {Man},
		\citenamefont {Xia},\ and\ \citenamefont {Franco}}]{man2018temperature}%
	\BibitemOpen
	\bibfield  {author} {\bibinfo {author} {\bibfnamefont {Z.-X.}\ \bibnamefont
			{Man}}, \bibinfo {author} {\bibfnamefont {Y.-J.}\ \bibnamefont {Xia}}, \ and\
		\bibinfo {author} {\bibfnamefont {R.~L.}\ \bibnamefont {Franco}},\
	}\href@noop {} {\bibfield  {journal} {\bibinfo  {journal} {Physical Review
				A}\ }\textbf {\bibinfo {volume} {97}},\ \bibinfo {pages} {062104} (\bibinfo
		{year} {2018})}\BibitemShut {NoStop}%
	\bibitem [{\citenamefont {{\c{C}}akmak}\ \emph {et~al.}(2017)\citenamefont
		{{\c{C}}akmak}, \citenamefont {Pezzutto}, \citenamefont {Paternostro},\ and\
		\citenamefont {M{\"u}stecapl{\i}o{\u{g}}lu}}]{ccakmak2017non}%
	\BibitemOpen
	\bibfield  {author} {\bibinfo {author} {\bibfnamefont {B.}~\bibnamefont
			{{\c{C}}akmak}}, \bibinfo {author} {\bibfnamefont {M.}~\bibnamefont
			{Pezzutto}}, \bibinfo {author} {\bibfnamefont {M.}~\bibnamefont
			{Paternostro}}, \ and\ \bibinfo {author} {\bibfnamefont {{\"O}.}~\bibnamefont
			{M{\"u}stecapl{\i}o{\u{g}}lu}},\ }\href@noop {} {\bibfield  {journal}
		{\bibinfo  {journal} {Physical Review A}\ }\textbf {\bibinfo {volume} {96}},\
		\bibinfo {pages} {022109} (\bibinfo {year} {2017})}\BibitemShut {NoStop}%
	\bibitem [{\citenamefont {Liu}\ \emph {et~al.}(2018)\citenamefont {Liu},
		\citenamefont {Zou},\ and\ \citenamefont {Fang}}]{yu2018quantum}%
	\BibitemOpen
	\bibfield  {author} {\bibinfo {author} {\bibfnamefont {Y.}~\bibnamefont
			{Liu}}, \bibinfo {author} {\bibfnamefont {H.-M.}\ \bibnamefont {Zou}}, \ and\
		\bibinfo {author} {\bibfnamefont {M.-F.}\ \bibnamefont {Fang}},\ }\href@noop
	{} {\bibfield  {journal} {\bibinfo  {journal} {arXiv preprint
				arXiv:1811.01154}\ } (\bibinfo {year} {2018})}\BibitemShut {NoStop}%
	\bibitem [{\citenamefont {Bhattacharya}\ \emph {et~al.}(2016)\citenamefont
		{Bhattacharya}, \citenamefont {Banerjee},\ and\ \citenamefont
		{Pati}}]{bhattacharya2016effect}%
	\BibitemOpen
	\bibfield  {author} {\bibinfo {author} {\bibfnamefont {S.}~\bibnamefont
			{Bhattacharya}}, \bibinfo {author} {\bibfnamefont {S.}~\bibnamefont
			{Banerjee}}, \ and\ \bibinfo {author} {\bibfnamefont {A.~K.}\ \bibnamefont
			{Pati}},\ }\href@noop {} {\bibfield  {journal} {\bibinfo  {journal} {Preprint
				at https://arxiv. org/abs/1601.04742}\ } (\bibinfo {year}
		{2016})}\BibitemShut {NoStop}%
	\bibitem [{\citenamefont {Chitambar}\ \emph {et~al.}(2016)\citenamefont
		{Chitambar}, \citenamefont {Streltsov}, \citenamefont {Rana}, \citenamefont
		{Bera}, \citenamefont {Adesso},\ and\ \citenamefont
		{Lewenstein}}]{AssistedDistillationofQuantumCoherence}%
	\BibitemOpen
	\bibfield  {author} {\bibinfo {author} {\bibfnamefont {E.}~\bibnamefont
			{Chitambar}}, \bibinfo {author} {\bibfnamefont {A.}~\bibnamefont
			{Streltsov}}, \bibinfo {author} {\bibfnamefont {S.}~\bibnamefont {Rana}},
		\bibinfo {author} {\bibfnamefont {M.~N.}\ \bibnamefont {Bera}}, \bibinfo
		{author} {\bibfnamefont {G.}~\bibnamefont {Adesso}}, \ and\ \bibinfo {author}
		{\bibfnamefont {M.}~\bibnamefont {Lewenstein}},\ }\href {\doibase
		10.1103/PhysRevLett.116.070402} {\bibfield  {journal} {\bibinfo  {journal}
			{Phys. Rev. Lett.}\ }\textbf {\bibinfo {volume} {116}},\ \bibinfo {pages}
		{070402} (\bibinfo {year} {2016})}\BibitemShut {NoStop}%
	\bibitem [{\citenamefont {Hu}\ and\ \citenamefont
		{Fan}(2016)}]{fanhuscireport2017}%
	\BibitemOpen
	\bibfield  {author} {\bibinfo {author} {\bibfnamefont {X.}~\bibnamefont
			{Hu}}\ and\ \bibinfo {author} {\bibfnamefont {H.}~\bibnamefont {Fan}},\
	}\href@noop {} {\bibfield  {journal} {\bibinfo  {journal} {Sci. Rep.}\
		}\textbf {\bibinfo {volume} {6}} (\bibinfo {year} {2016})}\BibitemShut
	{NoStop}%
	\bibitem [{\citenamefont {Hu}\ \emph {et~al.}(2016)\citenamefont {Hu},
		\citenamefont {Milne}, \citenamefont {Zhang},\ and\ \citenamefont
		{Fan}}]{hu2016quantum}%
	\BibitemOpen
	\bibfield  {author} {\bibinfo {author} {\bibfnamefont {X.}~\bibnamefont
			{Hu}}, \bibinfo {author} {\bibfnamefont {A.}~\bibnamefont {Milne}}, \bibinfo
		{author} {\bibfnamefont {B.}~\bibnamefont {Zhang}}, \ and\ \bibinfo {author}
		{\bibfnamefont {H.}~\bibnamefont {Fan}},\ }\href@noop {} {\bibfield
		{journal} {\bibinfo  {journal} {Sci. Rep.}\ }\textbf {\bibinfo {volume}
			{6}},\ \bibinfo {pages} {19365} (\bibinfo {year} {2016})}\BibitemShut
	{NoStop}%
	\bibitem [{\citenamefont {Liu}\ \emph {et~al.}(2011)\citenamefont {Liu},
		\citenamefont {Li}, \citenamefont {Huang}, \citenamefont {Li}, \citenamefont
		{Guo}, \citenamefont {Laine}, \citenamefont {Breuer},\ and\ \citenamefont
		{Piilo}}]{liu2011experimental}%
	\BibitemOpen
	\bibfield  {author} {\bibinfo {author} {\bibfnamefont {B.-H.}\ \bibnamefont
			{Liu}}, \bibinfo {author} {\bibfnamefont {L.}~\bibnamefont {Li}}, \bibinfo
		{author} {\bibfnamefont {Y.-F.}\ \bibnamefont {Huang}}, \bibinfo {author}
		{\bibfnamefont {C.-F.}\ \bibnamefont {Li}}, \bibinfo {author} {\bibfnamefont
			{G.-C.}\ \bibnamefont {Guo}}, \bibinfo {author} {\bibfnamefont {E.-M.}\
			\bibnamefont {Laine}}, \bibinfo {author} {\bibfnamefont {H.-P.}\ \bibnamefont
			{Breuer}}, \ and\ \bibinfo {author} {\bibfnamefont {J.}~\bibnamefont
			{Piilo}},\ }\href@noop {} {\bibfield  {journal} {\bibinfo  {journal} {Nat.
				Phys.}\ }\textbf {\bibinfo {volume} {7}},\ \bibinfo {pages} {931} (\bibinfo
		{year} {2011})}\BibitemShut {NoStop}%
	\bibitem [{\citenamefont {Wu}\ \emph {et~al.}(2017)\citenamefont {Wu},
		\citenamefont {Hou}, \citenamefont {Zhong}, \citenamefont {Yuan},
		\citenamefont {Xiang}, \citenamefont {Li},\ and\ \citenamefont
		{Guo}}]{wu2017experimentally}%
	\BibitemOpen
	\bibfield  {author} {\bibinfo {author} {\bibfnamefont {K.-D.}\ \bibnamefont
			{Wu}}, \bibinfo {author} {\bibfnamefont {Z.}~\bibnamefont {Hou}}, \bibinfo
		{author} {\bibfnamefont {H.-S.}\ \bibnamefont {Zhong}}, \bibinfo {author}
		{\bibfnamefont {Y.}~\bibnamefont {Yuan}}, \bibinfo {author} {\bibfnamefont
			{G.-Y.}\ \bibnamefont {Xiang}}, \bibinfo {author} {\bibfnamefont {C.-F.}\
			\bibnamefont {Li}}, \ and\ \bibinfo {author} {\bibfnamefont {G.-C.}\
			\bibnamefont {Guo}},\ }\href@noop {} {\bibfield  {journal} {\bibinfo
			{journal} {Optica}\ }\textbf {\bibinfo {volume} {4}},\ \bibinfo {pages} {454}
		(\bibinfo {year} {2017})}\BibitemShut {NoStop}%
	\bibitem [{\citenamefont {Vedral}(2002)}]{vedral2002role}%
	\BibitemOpen
	\bibfield  {author} {\bibinfo {author} {\bibfnamefont {V.}~\bibnamefont
			{Vedral}},\ }\href@noop {} {\bibfield  {journal} {\bibinfo  {journal} {Rev.
				Mod. Phys.}\ }\textbf {\bibinfo {volume} {74}},\ \bibinfo {pages} {197}
		(\bibinfo {year} {2002})}\BibitemShut {NoStop}%
	\bibitem [{\citenamefont {Streltsov}\ \emph
		{et~al.}(2017{\natexlab{b}})\citenamefont {Streltsov}, \citenamefont {Rana},
		\citenamefont {Bera},\ and\ \citenamefont
		{Lewenstein}}]{streltsov2017towards}%
	\BibitemOpen
	\bibfield  {author} {\bibinfo {author} {\bibfnamefont {A.}~\bibnamefont
			{Streltsov}}, \bibinfo {author} {\bibfnamefont {S.}~\bibnamefont {Rana}},
		\bibinfo {author} {\bibfnamefont {M.~N.}\ \bibnamefont {Bera}}, \ and\
		\bibinfo {author} {\bibfnamefont {M.}~\bibnamefont {Lewenstein}},\
	}\href@noop {} {\bibfield  {journal} {\bibinfo  {journal} {Phys. Rev. X}\
		}\textbf {\bibinfo {volume} {7}},\ \bibinfo {pages} {011024} (\bibinfo {year}
		{2017}{\natexlab{b}})}\BibitemShut {NoStop}%
	\bibitem [{\citenamefont {Streltsov}\ \emph
		{et~al.}(2015{\natexlab{b}})\citenamefont {Streltsov}, \citenamefont {Singh},
		\citenamefont {Dhar}, \citenamefont {Bera},\ and\ \citenamefont
		{Adesso}}]{streltsov2015measuring}%
	\BibitemOpen
	\bibfield  {author} {\bibinfo {author} {\bibfnamefont {A.}~\bibnamefont
			{Streltsov}}, \bibinfo {author} {\bibfnamefont {U.}~\bibnamefont {Singh}},
		\bibinfo {author} {\bibfnamefont {H.~S.}\ \bibnamefont {Dhar}}, \bibinfo
		{author} {\bibfnamefont {M.~N.}\ \bibnamefont {Bera}}, \ and\ \bibinfo
		{author} {\bibfnamefont {G.}~\bibnamefont {Adesso}},\ }\href@noop {}
	{\bibfield  {journal} {\bibinfo  {journal} {Phys. Rev. Lett}\ }\textbf
		{\bibinfo {volume} {115}},\ \bibinfo {pages} {020403} (\bibinfo {year}
		{2015}{\natexlab{b}})}\BibitemShut {NoStop}%
	\bibitem [{\citenamefont {Ma}\ \emph {et~al.}(2016)\citenamefont {Ma},
		\citenamefont {Yadin}, \citenamefont {Girolami}, \citenamefont {Vedral},\
		and\ \citenamefont {Gu}}]{ConvertingCoherencetoQuantumCorrelations}%
	\BibitemOpen
	\bibfield  {author} {\bibinfo {author} {\bibfnamefont {J.}~\bibnamefont
			{Ma}}, \bibinfo {author} {\bibfnamefont {B.}~\bibnamefont {Yadin}}, \bibinfo
		{author} {\bibfnamefont {D.}~\bibnamefont {Girolami}}, \bibinfo {author}
		{\bibfnamefont {V.}~\bibnamefont {Vedral}}, \ and\ \bibinfo {author}
		{\bibfnamefont {M.}~\bibnamefont {Gu}},\ }\href {\doibase
		10.1103/PhysRevLett.116.160407} {\bibfield  {journal} {\bibinfo  {journal}
			{Phys. Rev. Lett.}\ }\textbf {\bibinfo {volume} {116}},\ \bibinfo {pages}
		{160407} (\bibinfo {year} {2016})}\BibitemShut {NoStop}%
	\bibitem [{\citenamefont {Wu}\ \emph {et~al.}(2018)\citenamefont {Wu},
		\citenamefont {Hou}, \citenamefont {Zhao}, \citenamefont {Xiang},
		\citenamefont {Li}, \citenamefont {Guo}, \citenamefont {Ma}, \citenamefont
		{He}, \citenamefont {Thompson},\ and\ \citenamefont
		{Gu}}]{PRLresourceonversion}%
	\BibitemOpen
	\bibfield  {author} {\bibinfo {author} {\bibfnamefont {K.-D.}\ \bibnamefont
			{Wu}}, \bibinfo {author} {\bibfnamefont {Z.}~\bibnamefont {Hou}}, \bibinfo
		{author} {\bibfnamefont {Y.-Y.}\ \bibnamefont {Zhao}}, \bibinfo {author}
		{\bibfnamefont {G.-Y.}\ \bibnamefont {Xiang}}, \bibinfo {author}
		{\bibfnamefont {C.-F.}\ \bibnamefont {Li}}, \bibinfo {author} {\bibfnamefont
			{G.-C.}\ \bibnamefont {Guo}}, \bibinfo {author} {\bibfnamefont
			{J.}~\bibnamefont {Ma}}, \bibinfo {author} {\bibfnamefont {Q.-Y.}\
			\bibnamefont {He}}, \bibinfo {author} {\bibfnamefont {J.}~\bibnamefont
			{Thompson}}, \ and\ \bibinfo {author} {\bibfnamefont {M.}~\bibnamefont
			{Gu}},\ }\href {\doibase 10.1103/PhysRevLett.121.050401} {\bibfield
		{journal} {\bibinfo  {journal} {Phys. Rev. Lett.}\ }\textbf {\bibinfo
			{volume} {121}},\ \bibinfo {pages} {050401} (\bibinfo {year}
		{2018})}\BibitemShut {NoStop}%
	\bibitem [{\citenamefont {Raussendorf}\ and\ \citenamefont
		{Briegel}(2001)}]{RausB01}%
	\BibitemOpen
	\bibfield  {author} {\bibinfo {author} {\bibfnamefont {R.}~\bibnamefont
			{Raussendorf}}\ and\ \bibinfo {author} {\bibfnamefont {H.~J.}\ \bibnamefont
			{Briegel}},\ }\href@noop {} {\bibfield  {journal} {\bibinfo  {journal} {Phys.
				Rev. Lett.}\ }\textbf {\bibinfo {volume} {86}},\ \bibinfo {pages} {5188}
		(\bibinfo {year} {2001})}\BibitemShut {NoStop}%
	\bibitem [{\citenamefont {Vedral}\ and\ \citenamefont
		{Plenio}(1998)}]{VedrP98}%
	\BibitemOpen
	\bibfield  {author} {\bibinfo {author} {\bibfnamefont {V.}~\bibnamefont
			{Vedral}}\ and\ \bibinfo {author} {\bibfnamefont {M.~B.}\ \bibnamefont
			{Plenio}},\ }\href@noop {} {\bibfield  {journal} {\bibinfo  {journal} {Phys.
				Rev. A}\ }\textbf {\bibinfo {volume} {57}},\ \bibinfo {pages} {1619}
		(\bibinfo {year} {1998})}\BibitemShut {NoStop}%
	\bibitem [{\citenamefont {Ollivier}\ and\ \citenamefont
		{Zurek}(2001)}]{ollivier2001quantum}%
	\BibitemOpen
	\bibfield  {author} {\bibinfo {author} {\bibfnamefont {H.}~\bibnamefont
			{Ollivier}}\ and\ \bibinfo {author} {\bibfnamefont {W.~H.}\ \bibnamefont
			{Zurek}},\ }\href {\doibase 10.1103/PhysRevLett.88.017901} {\bibfield
		{journal} {\bibinfo  {journal} {Phys. Rev. Lett.}\ }\textbf {\bibinfo
			{volume} {88}},\ \bibinfo {pages} {017901} (\bibinfo {year}
		{2001})}\BibitemShut {NoStop}%
	\bibitem [{\citenamefont {Henderson}\ and\ \citenamefont
		{Vedral}(2001)}]{henderson2001classical}%
	\BibitemOpen
	\bibfield  {author} {\bibinfo {author} {\bibfnamefont {L.}~\bibnamefont
			{Henderson}}\ and\ \bibinfo {author} {\bibfnamefont {V.}~\bibnamefont
			{Vedral}},\ }\href@noop {} {\bibfield  {journal} {\bibinfo  {journal} {J.
				Phys. A}\ }\textbf {\bibinfo {volume} {34}},\ \bibinfo {pages} {6899}
		(\bibinfo {year} {2001})}\BibitemShut {NoStop}%
	\bibitem [{\citenamefont {Modi}\ \emph {et~al.}(2012)\citenamefont {Modi},
		\citenamefont {Brodutch}, \citenamefont {Cable}, \citenamefont {Paterek},\
		and\ \citenamefont {Vedral}}]{RevModPhysdiscord}%
	\BibitemOpen
	\bibfield  {author} {\bibinfo {author} {\bibfnamefont {K.}~\bibnamefont
			{Modi}}, \bibinfo {author} {\bibfnamefont {A.}~\bibnamefont {Brodutch}},
		\bibinfo {author} {\bibfnamefont {H.}~\bibnamefont {Cable}}, \bibinfo
		{author} {\bibfnamefont {T.}~\bibnamefont {Paterek}}, \ and\ \bibinfo
		{author} {\bibfnamefont {V.}~\bibnamefont {Vedral}},\ }\href {\doibase
		10.1103/RevModPhys.84.1655} {\bibfield  {journal} {\bibinfo  {journal} {Rev.
				Mod. Phys.}\ }\textbf {\bibinfo {volume} {84}},\ \bibinfo {pages} {1655}
		(\bibinfo {year} {2012})}\BibitemShut {NoStop}%
\end{thebibliography}
\end{document}